\documentclass[fleqn,usenatbib]{mnras}

\usepackage{newtxtext,newtxmath}
\usepackage{wasysym}
\usepackage{graphicx}

\usepackage[T1]{fontenc}

\DeclareRobustCommand{\VAN}[3]{#2}
\let\VANthebibliography\thebibliography
\def\thebibliography{\DeclareRobustCommand{\VAN}[3]{##3}\VANthebibliography}

\usepackage{graphicx}	
\usepackage{amsmath}	

\usepackage{orcidlink}

\title[Not So Fast Kepler-1513]{Not So Fast Kepler-1513: \newline A Perturbing Planetary Interloper in the Exomoon Corridor}

\author[D. A. Yahalomi et al.]{Daniel A. Yahalomi \orcidlink{0000-0003-4755-584X},$^{1}$\thanks{E-mail: daniel.yahalomi@columbia.edu}\thanks{LSSTC DSFP Fellow}
David Kipping \orcidlink{0000-0002-4365-7366},$^{1}$
David Nesvorn\'y \orcidlink{0000-0002-4547-4301},$^{2}$
Paul A. Dalba \orcidlink{0000-0002-4297-5506},$^{3, 4, 5}$
\newauthor
Paul Benni \orcidlink{0000-0001-6981-8722},$^{6}$
Ceiligh Cacho-Negrete,$^{7}$
Karen Collins \orcidlink{0000-0001-6588-9574},$^{8}$
Joel T. Earwicker \orcidlink{0000-0001-5915-1127},$^{4, 9}$
\newauthor
John Arban Lewis \orcidlink{0000-0001-5199-3522},$^{8}$
Kim K. McLeod \orcidlink{0000-0001-9504-1486},$^{7}$
Richard P. Schwarz \orcidlink{0000-0001-8227-1020},$^{8}$
and Gavin Wang \orcidlink{0000-0003-3092-4418}$^{10}$
\\
$^{1}$Department of Astronomy, Columbia University, 550 W 120th St., New York, NY 10027, USA\\
$^{2}$Southwest Research Institute, 1050 Walnut St, Suite 300, Boulder, CO 80302, USA\\
$^{3}$Heising-Simons 51 Pegasi b Postdoctoral Fellow\\
$^{4}$Department of Astronomy and Astrophysics, University of California, Santa Cruz, CA 95064, USA\\
$^{5}$SETI Institute, Carl Sagan Center, 339 Bernardo Ave, Suite 200, Mountain View, CA 94043, USA\\
$^{6}$Acton Sky Portal private observatory, Acton, MA, USA\\
$^{7}$Department of Astronomy, Wellesley College, Wellesley, MA 02481, USA\\
$^{8}$Center for Astrophysics \textbar \ Harvard \& Smithsonian, 60 Garden Street, Cambridge, MA 02138, USA\\
$^{9}$Boyce Research Initiatives and Education Foundation, 3540 Carleton St., San Diego, CA
92106, USA\\
$^{10}$Department of Physics \& Astronomy, Johns Hopkins University, 3400 N. Charles Street, Baltimore, MD 21218, USA\\
}

\date{Accepted XXX. Received YYY; in original form ZZZ}

\pubyear{2023}

\begin{document}
\label{firstpage}
\pagerange{\pageref{firstpage}--\pageref{lastpage}}
\maketitle

\begin{abstract}
Transit Timing Variations (TTVs) can be induced by a range of physical phenomena, including planet-planet interactions,  planet-moon interactions, and stellar activity. Recent work has shown that roughly half of moons would induce fast TTVs with a short period in the range of two-to-four orbits of its host planet around the star. An investigation of the \textit{Kepler} TTV data in this period range identified one primary target of interest, Kepler-1513\,b. Kepler-1513\,b is a $8.05^{+0.58}_{-0.40}$\,$R_\oplus$ planet orbiting a late G-type dwarf at $0.53^{+0.04}_{-0.03}$\,AU. Using \textit{Kepler} photometry, this initial analysis showed that Kepler-1513\,b's TTVs were consistent with a moon. Here, we report photometric observations of two additional transits nearly a decade after the last \textit{Kepler} transit using both ground-based observations and space-based photometry with TESS. These new transit observations introduce a previously undetected long period TTV, in addition to the original short period TTV signal. Using the complete transit dataset, we investigate whether a non-transiting planet, a moon, or stellar activity could induce the observed TTVs. We find that only a non-transiting perturbing planet can reproduce the observed TTVs. We additionally perform transit origami on the \textit{Kepler} photometry, which independently applies pressure against a moon hypothesis. Specifically, we find that Kepler-1513\,b's TTVs are consistent with an exterior non-transiting $\sim$Saturn mass planet, Kepler-1513\,c, on a wide orbit, $\sim$5$\%$ outside a 5:1 period ratio with Kepler-1513\,b. This example introduces a previously unidentified cause for planetary interlopers in the exomoon corridor, namely an insufficient baseline of observations.
\end{abstract}




\section{Introduction} \label{sec:intro}

\begin{figure*}
    \centering 
    \includegraphics[width=.8\textwidth]{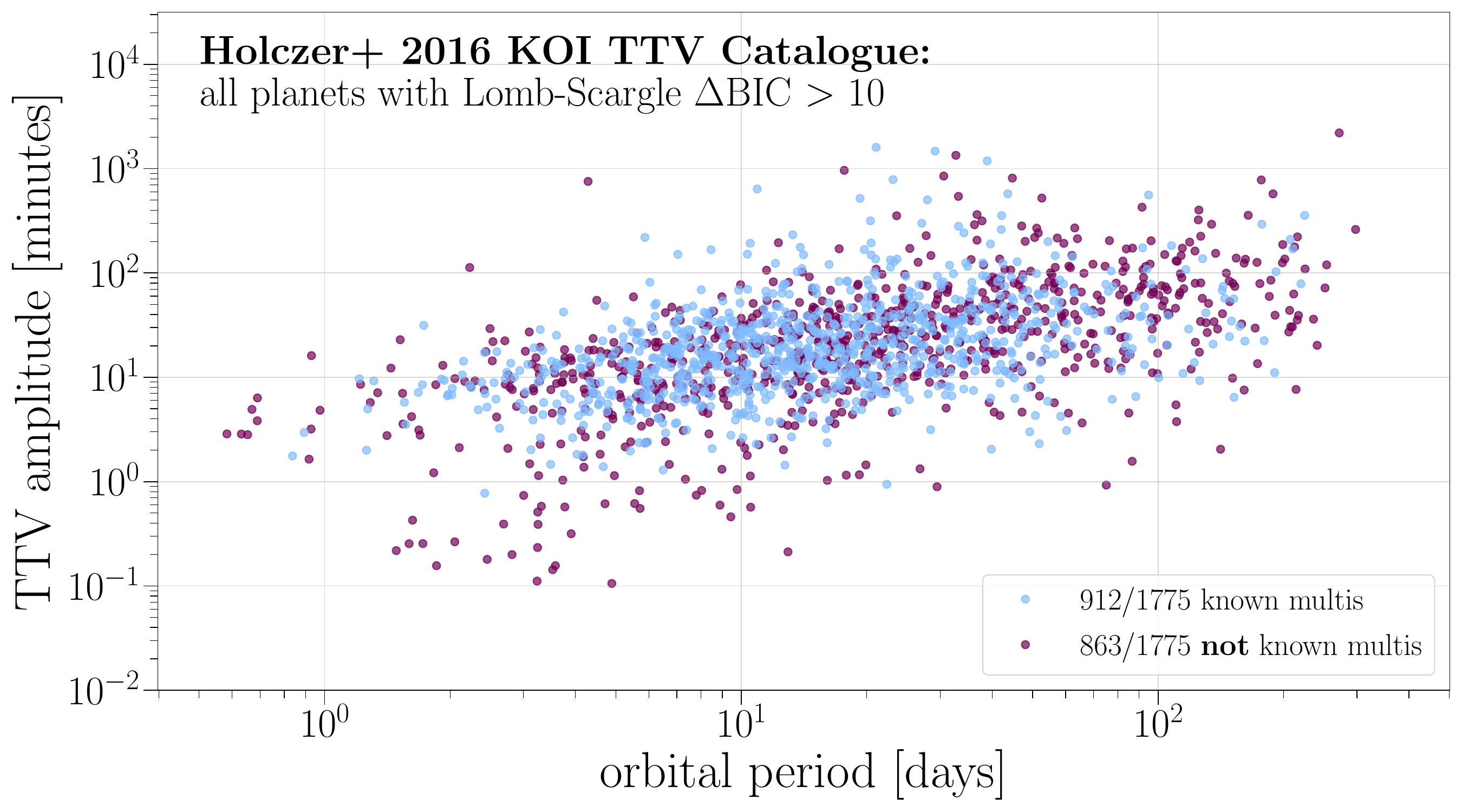}
    \caption{All systems in \citet{Holczer2016} \textit{Kepler} TTV catalog for which the TTV signal passes a threshold for its Lomb-Scargle period of $\Delta$BIC $>$ 10 (1775 out of 2599 in total). Here, $\Delta$BIC is defined as BIC$_{null}$ - BIC$_{LS}$. In light blue are all systems that are known multi-planet systems (912 systems). In magenta are the rest of the planets in the catalog, and thus are not known multi-planet systems (863 systems). This shows the need to develop TTV model selection methods to differentiate between possible causes.} 
    \label{fig: holczer_catalogue}
\end{figure*}

\subsection{Background}
Transit Timing Variations (TTVs) are discrepancies between the observed transit time of a transiting exoplanet and that of a Keplerian linear ephemeris. The importance of transit timing variations for exoplanet study was first discussed in \citet{Dobrovolskis1996} and \citet{Miralda-Escude2002}.  Subsequently, \citet{Holman2005} and \citet{Agol2005} pointed out the importance of orbits near mean motion resonance (MMR) planets for TTV studies. For TTV signals caused by near MMR orbits, there exists a degeneracy between the eccentricity and the mass ratios between the planets and the host star. This degeneracy can be broken using chopping signals, second order harmonics that are often caused when planets pass each other in conjunction at the synodic period \citep{Lithwick2012, Nesvorny2014, Schmitt2014, Deck2015}. \citet{AgolFabrycky2018} provide an in depth discussion of planet-planet TTVs. 

In general, TTVs are observational manifestations of a slightly varying period of the observed exoplanet due to another gravitational force in the system, with the caveat that stellar activity can also mimic TTV signals and thus cause false positives \citep{Sanchis-Ojeda2011, Mazeh2013, Szabo2013, Oshagh2013, Holczer2015, Mazeh2015}. TTV signals of hundreds of \textit{Kepler} planets have been catalogued \citep{Lissauer2011, Mazeh2013, Holczer2016, Ofir2018} and dozens of TTV systems have been confirmed via their planet-planet interactions (eg. \citet{Holman2010}, \citet{Ballard2011}, \citet{Nesvorny2012}, \citet{Nesvorny2013}, \citet{Hadden2014}, \citet{Pearson2019}, \citet{Jontof-Hutter2019}, \citet{Trifonov2021}). 

However, many of the systems with identified TTVs have an undetermined cause, be it planet-planet induced TTVs, planet-moon induced TTVs, or stellar activity induced TTVs (see Figure \ref{fig: holczer_catalogue}). Additionally, it is challenging to determine the planetary properties for a set of observed TTVs assuming a planet-planet TTV \citep{Jontof-Hutter2016}. Dynamical \textit{N}-body simulations \citep[\texttt{SWIFT},][]{Levison1994, Nesvorny2013}, \citep[\texttt{TTVFast},][]{Deck2014} use planet parameters to compute a set of predicted transit times. However, in practice, inverting the problem becomes non-trivial as comparing these dynamically modelled transit times to a set of observed transit times is a  non-linear model, that often has multi-modal solutions, and with parameters that are highly correlated \citep{Tuchow2019}.

As such, performing parameter estimation of an observed TTV is difficult even when one assumes that the TTV is caused by a planet-planet interaction. Widening the frame of possibilities to non planet-planet induced TTVs only makes the question more complicated. Of particular note to this paper is the study of moon induced TTVs.

Moons are prevalent in our Solar System and appear to be a natural byproduct of planet formation \citep{Crida2012}. A hypothetical moon orbiting an exoplanet (henceforth exomoon) would gravitationally pull on its host planet with a period of the exomoon's orbital period. No Solar-System analogous exomoons have been detected to date. However, as these satellites are quite small this can be explained by detection limitations of our current instrumentation \citep{Kipping2012, Cassese2022}. Recently, two exomoon candidates have emerged: Kepler-1625b-i \citep{Teachey2018} and Kepler-1708b-i \citep{Kipping2022}. These exomoon candidates are larger than the Earth, with radii of $\sim$5 and $\sim$2.5 Earth radii respectively -- defying most current models of moon formation \citep{Canup2006, Cilibrasi2018}. Kepler-1625b-i exhibits TTVs that were robustly recovered by a series of independent studies, however, the exomoon transit was recovered by one independent study \citep{Heller2019} but was not recovered by another \citep{Kreidberg2019}. Kepler-1708b-i exhibits a transit-like signal, but its TTVs cannot be studied as there are only two transit observations.

In order to search for exomoon candidates via TTV signals, recently, a statistical approach has been proposed via investigation of high frequency (or short period) TTVs, the so-called ``exomoon corridor'' \citep{Kipping2021}. The argument for the exomoon corridor goes as follows: exomoon induced TTVs will always be undersampled as an exomoon's orbital period will be shorter than its host planet's orbital period. Thus, with a maximum of one transit observation per cycle, the exomoon TTV signal will manifest as an alias of the true period. These aliases are much more likely to manifest near the Nyquist rate, which is the highest frequency that can be used at a given sampling rate in order to fully reconstruct a signal. It turns out that $\sim$50$\%$ of the aliases will occur in a period range between two-to-four planetary orbital cycles independent of physical and orbital assumptions for the moon -- thus defining the so-called ``exomoon corridor,'' and identifying an area in TTV space of particular interest in the search for exomoon induced TTVs.

\begin{figure*}
\centering
\includegraphics[width=0.49\textwidth]{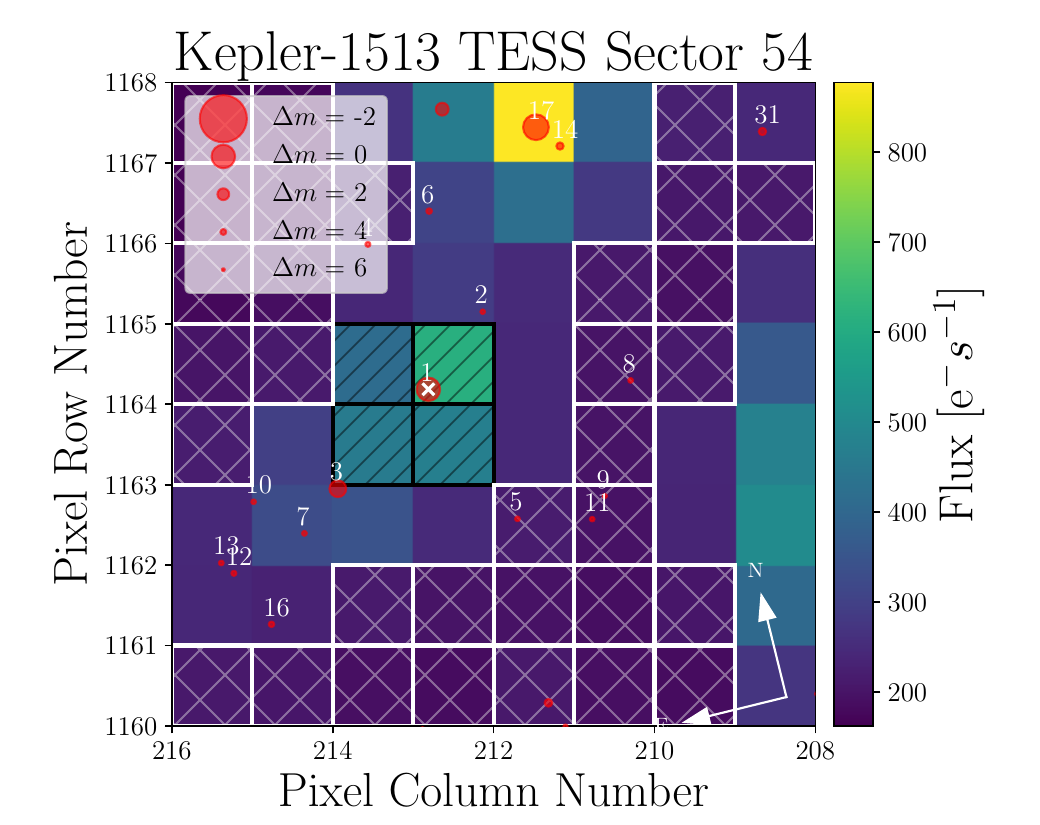}
\hfill
\includegraphics[width=0.49\textwidth]{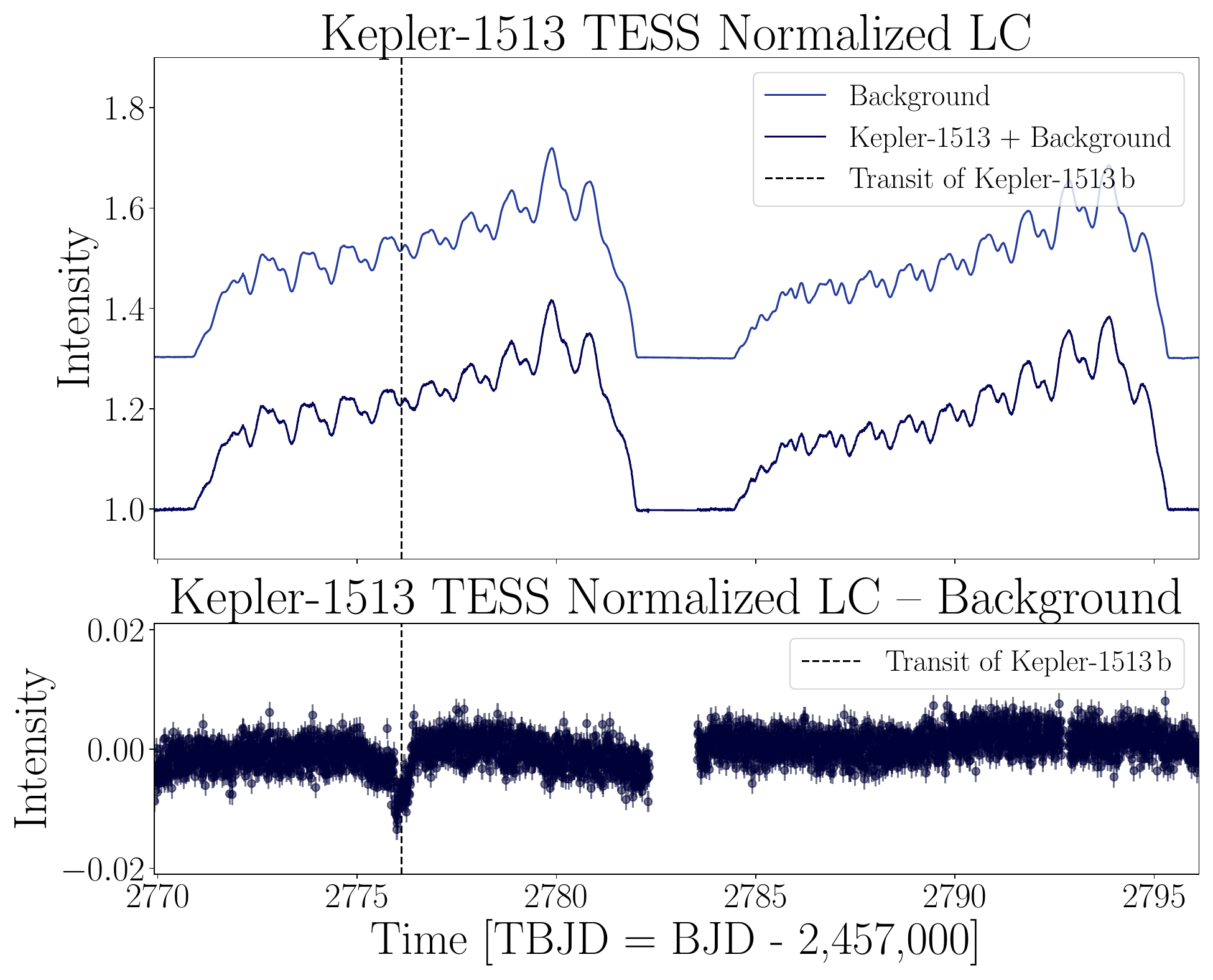}
\caption{\textbf{Left:} TESS aperture used for Sector 54 observation of Kepler-1513. In black is the selected target aperture and in white is the selected background aperture. The red dots represent neighboring stars from \textit{Gaia} DR3. Plot generated with a customized version of \texttt{tpfplotter}. \textbf{Right:} TESS undetrended light curve (LC) for Sector 54, extrapolated from the target and background apertures. The top half shows the flux from the background aperture and the flux from the background aperture and the target aperture. The bottom shows the background subtracted and normalized flux of the target aperture. The vertical dashed line shows Kepler-1513\,b's expected transit using the \textit{Kepler} linear ephemeris.}
\label{fig: tess_data}
\end{figure*}

\citet{KippingYahalomi2022} investigated the full \citet{Holczer2016} \textit{Kepler} TTV catalog, which, starting with an initial data set of 2599 KOIs, led to eleven candidate exomoon corridor TTVs. These eleven candidate TTV systems were then run through three additional statistical tests: (1) tested if the TTVs are statisticially significant by comparing the (Bayesian Inference Criteria) BIC of a linear ephemeris with that of the TTV model, (2) tested if the TTVs are statistically significant periodically by holding out 20$\%$ of the dataset and seeing if the TTV model is still statistically better than the linear ephemeris model, and (3) tested if the observations support a statistically significant non-zero moon mass. Only one of these planetary systems survived all three of these tests: Kepler-1513\,b. Using \texttt{forecaster} \citep{ChenKipping2017}, \citet{KippingYahalomi2022} estimated a mass for Kepler-1513\,b of $48^{+35}_{-21}$\,M$_\oplus$ and a maximum moon mass via stability of $4.4^{+13.4}_{-3.5}$\,M$_\oplus$. Subsequently, \citet{KisareFabrycky2023} showed that the tidal dissipation of Kepler-1513\,b due to a hypothetical satellite, with the satellite mass estimated from the amplitude of the observed \textit{Kepler} TTVs, was not unreasonable for a stable planet-moon system.

Despite the substantial effort that has been put into identifying this object, to date there has been neither a N-body analysis of the physical plausibility of the planet-planet hypothesis via TTV inversion nor a full photodynamical analysis of the planet-moon hypothesis. Here, we present two additional transit observations, nearly a decade after the last \textit{Kepler} transit that introduces an additional previously undetectable long period TTV. We then present a photodynamical model for planet-moon induced TTVs and an N-body model for planet-planet induced TTVs, demonstrate that the TTVs are not consistent with stellar activity induced signals, and argue that the Kepler-1513\,b TTV data favors a planet-planet system, which previously hid in the exomoon corridor due to insufficient baseline of observations.

\section{Methods}

\subsection{Photometry}

\begin{figure*}
\centering 

\includegraphics[width=\textwidth]{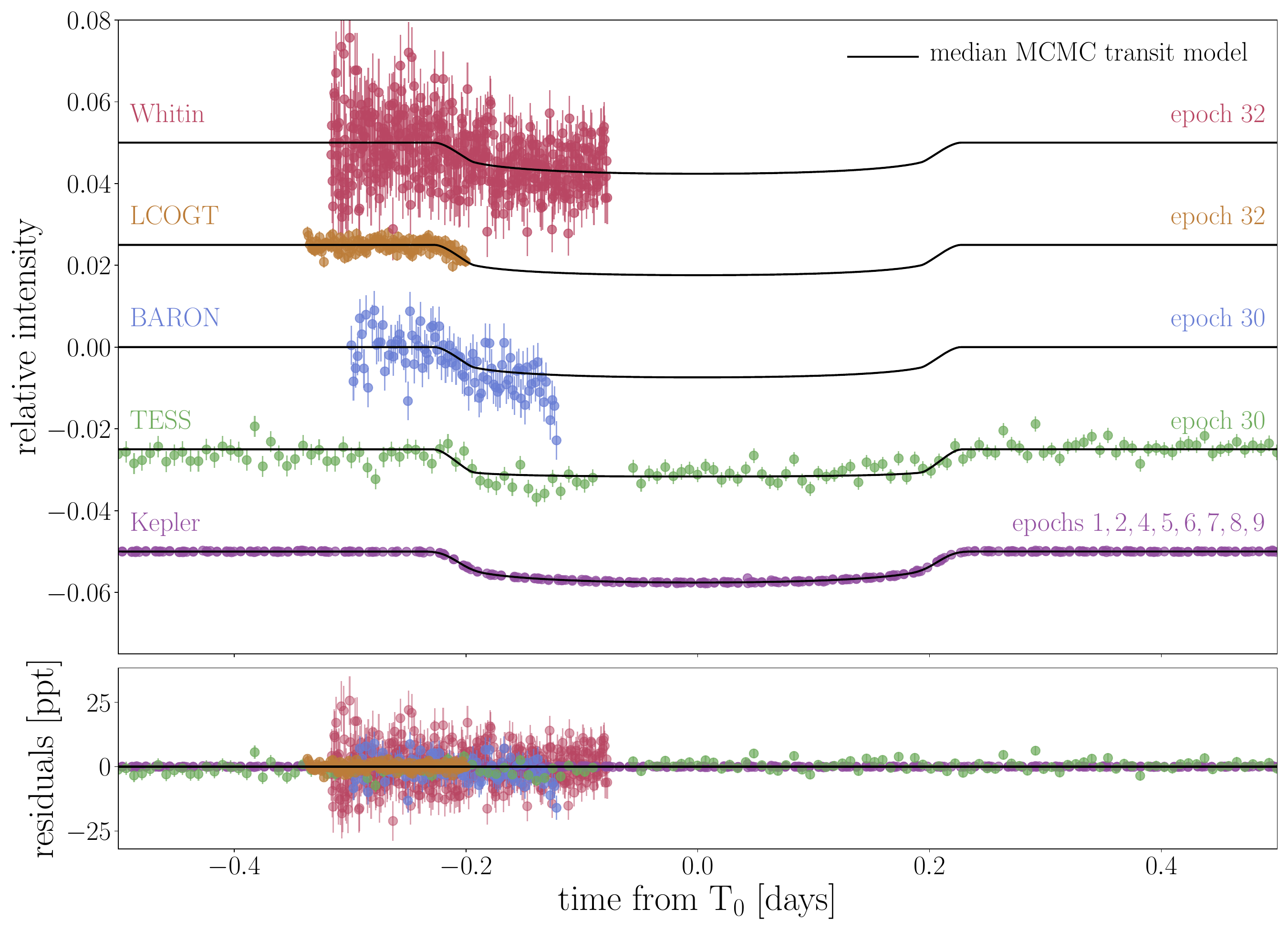}
\caption{Phase folded light curve using all ten observed transits of Kepler-1513\,b with \textit{Kepler}, TESS, BARON, LCOGT, and Whitin data.}
\label{fig: phase_fold_all}

\end{figure*}

\subsubsection{\textit{Kepler}}

\textit{Kepler} observed nine transits of Kepler-1513\,b from 2009 to 2013. The third transit observed ($\sim$2,455,432 BJD) occurred near a large systematic event and only a partial transit was observed, so the data was not included in the dynamical modeling that follows -- resulting in eight \textit{Kepler} transits in total.

\subsubsection{TESS}

We were able to extract a transit of Kepler-1513\,b on July 15, 2022 from the TESS pixel level data, as no light curve (LC) was released by the TESS team for this transit. As presented in \citet{Jontof-Hutter2022}, TESS is powerful in its ability to obtain follow-up observations of \textit{Kepler} TTV systems, as TESS' 27-day sectors can allow for full transit observations when the long transit durations that are common for \textit{Kepler} TTV systems are too long to be observed with ground based observations. Long transit durations are common for \textit{Kepler} TTVs as the signal-to-noise ratio of TTVs increases with orbital period, which biases planetary mass detections to longer orbital periods, and thus longer transit durations \citep{Steffen2016}. In order to recover the TESS LC, we used the \texttt{lightkurve} function \texttt{TESScut} to load the TESS sector 54 data for Kepler-1513. We then chose an aperture that contained the light from the target star, using the \texttt{lightkurve} function \texttt{create\_threshold\_mask} with a threshold cut of three and a reference pixel as the center. Next, we determined the background flux, again using the \texttt{lightkurve} function \texttt{create\_threshold\_mask}, but this time with a threshold cut of 0.001 and no reference pixel. This allows us to identify and subtract out the background flux. We test the quality of our aperture selection by cross-referencing the aperture with the location of neighboring stars from \textit{Gaia} Data Release 3 using \texttt{tpfplotter}, and while there is a star near the edge of the aperture (bottom left pixel of the \texttt{tpfplotter} plot on the left in Figure \ref{fig: tess_data}), we chose not to shrink the aperture as four pixels is already quite small. In order to account for the effect of this nearby star, we added a blend factor into the transit model for the TESS data, which is later described in more detail. The selected aperture, background aperture, and the flux contributions from neighboring stars can be seen in Figure \ref{fig: tess_data}.

\subsubsection{BARON}

We observed the ingress of an additional transit of Kepler-1513\,b on July 15, 2022 using a Planewave CDK17 (0.43m) telescope located at the Boyce-Astro Research Observatory-North (BARON) at Sierra Remote Observatory (SRO). BARON observations were made with a QHY 600 CMOS camera and the Sloan $i'$-band. The raw frames were corrected for bias and dark current before being flattened with a median combined flat field image. Aperture photometry of Kepler-1513 proceeded following the techniques of \citet{Dalba2017}. Briefly, we used floating apertures to measure the flux and background signal of Kepler-1513 and all reasonably bright sources in the images. The flux of Kepler-1513 was normalized to hundreds of reference light curves generated by averaging the time series flux of various combinations of the background sources. The final light curve was chosen as the one which minimized the photometric scatter in the pre-transit baseline. We repeated this procedure with several different photometric aperture sizes to further minimize out-of-transit scatter. The light curve was detrended using a linear fit to airmass.

\subsubsection{LCOGT}

We observed a short ingress window of Kepler-1513\,b on 2023 June 02 UTC in the Sloan $i'$-band using the Las Cumbres Observatory Global Telescope \citep[LCOGT;][]{Brown2013} 1.0\,m network node at Teide Observatory on the island of Tenerife (TEID). The images were calibrated by the standard LCOGT {\tt BANZAI} pipeline \citep{McCully2018} and differential photometric data were extracted using {\tt AstroImageJ} \citep{Collins2017}. We used circular photometric apertures with radius $2\farcs0$ centered on Kepler-1513, which avoided most of the flux from the $2\farcs6$ neighbor TIC 1716106681. We selected nine comparison stars that minimized transit model residuals. The best zero, one, or two parametric detrend vectors were retained if joint linear fits to them plus a transit model decreased the BIC for the fit by at least two per detrend parameter. We found that the pair of vectors full-Width-half-maximum (FWHM) and sky-background of the target star provided the best improvement to the light curve fit while also being justified by improved BIC values.

\subsubsection{Whitin}

We observed an ingress window of Kepler-1513\,b in the Cousins $R$-band on 2023 June 02 UTC using the Whitin observatory 0.7\,m telescope in Wellesley, MA. The $1024\times1024$ SBIG STL-1001E detector has an image scale $1\farcs08$pixel$^{-1}$, resulting in a $18\farcm2\times18\farcm2$ field of view. The image data were calibrated and photometric data were extracted using {\tt AstroImageJ}. We used a circular photometric aperture with radius $7\farcs6$ centered on Kepler-1513, which included all of the flux from the neighbor $2\farcs6$ south. However, the neighbor is more than $\sim5$ magnitudes fainter, so its affect on the light curve depth would be negligible. The light curve was detrended using a linear fit to airmass.

\subsection{Democratic Detrending}

\begin{table*}
\centering
\renewcommand{\arraystretch}{1.3}  

\caption{Parameters for the Kepler-1513\,b transit model. Transit modeled using a classic Mandel-Agol analytic light curve model \citep{MandelAgol2002} with \texttt{exoplanet} \citep{Foreman-Mackey2021} and PyMC3 \citep{Salvatier2016} gradient based MCMC exploration of the parameter space. Credible interval derived from the [16$\%$, 50$\%$, \& 84$\%$] quantiles of the \textit{a-posteriori} solution.}
\label{table: transit_params}
\begin{tabular}{ c  c  c  c }
\hline
Parameter & Definition & Prior & Credible Intervals \\
\hline 
\multicolumn{4}{c}{\centering \textbf{MCMC Model Parameters}} \\

$R_P / R_*$ & Ratio-of-Radii & $\mathcal{U}$(0, 1)  & $0.07485^{+0.00031}_{-0.00019}$\\

$\rho_*$ [$g\,cm^{-3}$] & Stellar Density & $\log_e \mathcal{U}$(10$^{-3}$, 10$^3$) &  $1.304^{+0.019}_{-0.047}$  \\

b & Impact Parameter & $\mathcal{U}$(0, 2) & $0.106^{+0.087}_{-0.072}$\\

$q_{1, \textrm{\textit{Kepler}}}$ & \textit{Kepler} Limb-Darkening Parameter &  $\mathcal{U}$(0, 1) & $0.451^{+0.041}_{-0.038}$ \\

$q_{2, \textrm{\textit{Kepler}}}$ & \textit{Kepler} Limb-Darkening Parameter & $\mathcal{U}$(0, 1) & $0.323^{+0.030}_{-0.027}$ \\

$q_{1, \textrm{TESS}}$ & TESS Limb-Darkening Parameter & $\mathcal{U}$(0, 1) & $0.09^{+0.15}_{-0.06}$
 \\

$q_{2, \textrm{TESS}}$ & TESS Limb-Darkening Parameter & $\mathcal{U}$(0, 1) & $0.20^{+0.34}_{-0.16}$ \\

$\mathcal{B}_\textrm{TESS}$ & TESS Blend Factor & $\log_e \mathcal{U}$(1, 10)  & $1.000055^{0.000031}_{-0.000029}$ \\

$\tau_1$ [KBJD] &  Time of Transit Minimum  & $\mathcal{U}$(276.5, 278.5) & $277.50563^{+0.0004}_{-0.00038}$ \\

$\tau_2$ [KBJD] &  Time of Transit Minimum  & $\mathcal{U}$(437.4, 439.4) & $438.38777^{+0.00043}_{-0.00045}$ \\

$\tau_4$ [KBJD] &  Time of Transit Minimum  & $\mathcal{U}$(759.2, 761.2) & $760.15460^{+0.00046}_{-0.00046}$\\

$\tau_5$ [KBJD] &  Time of Transit Minimum  & $\mathcal{U}$(920.0, 922.0) & $921.04575^{+0.00041}_{-0.00039}$ \\

$\tau_6$ [KBJD] &  Time of Transit Minimum  & $\mathcal{U}$(1080.9, 1082.9) & $1081.92944^{+0.0004}_{-0.00041}$ \\

$\tau_7$ [KBJD] &  Time of Transit Minimum  & $\mathcal{U}$(1241.8, 1243.8) & $1242.80899^{+0.00035}_{-0.00035}$ \\

$\tau_8$ [KBJD] &  Time of Transit Minimum  & $\mathcal{U}$(1402.7, 1404.7) & $1403.70186^{+0.00043}_{-0.00044}$ \\

$\tau_9$ [KBJD] &  Time of Transit Minimum  & $\mathcal{U}$(1563.6, 1565.6) & $1564.57761^{+0.00038}_{-0.00035}$ \\

$\tau_{30}$ [KBJD] &  Time of Transit Minimum  & $\mathcal{U}$(4942.1, 4944.1) & $4943.1384^{+0.0033}_{-0.0034}$ \\

$\tau_{32}$ [KBJD] &  Time of Transit Minimum  & $\mathcal{U}$(5263.9, 5265.9) & $5264.918^{+0.0017}_{-0.0015}$ \\

\hline
\multicolumn{4}{c}{\centering \textbf{Assumed Limb-Darkening Parameters for Partial Transits from \citet{Claret2011}}} \\
$q_{1, \textrm{BARON}}$ & BARON Limb-Darkening Parameter &  ... & 0.370 \\

$q_{2, \textrm{BARON}}$ & BARON Limb-Darkening Parameter &  ...  & 0.313 \\

$q_{1, \textrm{LCOGT}}$ & LCOGT Limb-Darkening Parameter &  ...  & 0.370 \\

$q_{2, \textrm{LCOGT}}$ & LCOGT Limb-Darkening Parameter &  ...  & 0.313 \\

$q_{1, \textrm{Whitin}}$ & Whitin Limb-Darkening Parameter &   ...  & 0.4529 \\

$q_{2, \textrm{Whitin}}$ & Whitin Limb-Darkening Parameter &  ...  & 0.3342 \\

\hline
\multicolumn{4}{c}{\centering \textbf{Infered Parameters}} \\
P [days] & Linear Ephemeris Orbital Period  & ... & $160.884088^{0.000066}_{-0.000068}$\\

t$_0$  [KBJD] & Linear Ephemeris Time of First Transit Minimum & ... &  $277.50639^{+0.00032}_{-0.00030}$ \\

$R_P$  [R$_\oplus$] & Radius of Kepler-1513\,b  & ... &  $8.594^{+0.036}_{-0.022}$\\

\hline
\end{tabular}
\end{table*}

In order to obtain accurate transit times, first we must detrend the 10 Kepler-1513\,b transits, removing as much stellar activity as possible. 

For the space-based photometry, from \textit{Kepler} and TESS, we developed an open source Python code package, called the \texttt{democratic\_detrender}, that implements democratic (also previously called method marginalized) detrending. To summarize, democratic detrending uses several different detrending algorithms, and then takes the median solution between the detrending methods for each data point. In our work, we used four distinct detrending algorithms, which have each been shown in the literature to be efficient and accurate models for stellar noise: \texttt{CoFiAM}, \texttt{polyAM}, \texttt{local}, and \texttt{GP}. For a more detailed description of each detrending algorithm, see Appendix \ref{sec: democratic_detrender} and for a similar application see \citet{Kipping2022}. The \textit{Kepler} and TESS detrended LCs can be seen in Figure \ref{fig: detrended_lc}. This democratic detrending package is accessible via GitHub.\footnote{\href{https://github.com/dyahalomi/democratic\_detrender}{https://github.com/dyahalomi/democratic\_detrender}}

\section{Analysis and Results}

\subsection{Modeling the Transits of Kepler-1513\hspace{0.15em}b}
We now need to fit a transit model and determine the transit times. In order to do so, we set up a classic Mandel-Agol analytic light curve (LC) model \citep{MandelAgol2002} using the \texttt{exoplanet} toolkit \citep{Foreman-Mackey2021}. \texttt{exoplanet} uses PyMC3, to perform No-U-Turn Sampling, a variant of Hamiltonian Monte Carlo that automatically adapts the step size and trajectory length during sampling, to sample a posterior \citep{Salvatier2016}. 

As we had photometry from five different instruments, we modeled each instrument with unique limb-darkening parameters. For the ground-based observations, which only observed partial transits, we fixed the limb-darkening parameters to the corresponding values for the stars mass, radius, and observational filter from the results presented in \citet{Claret2011}. The limb-darkening values used can be seen in Table \ref{table: transit_params}. As previously mentioned, we additionally included a blend factor, $\mathcal{B}_\textrm{TESS}$, for the TESS data to account for contamination from a nearby star whose light may have contributed to the light observed by one of the pixels in the selected aperture. The blend factor is defined as the ratio of the total flux to that of the target star flux:
\begin{equation}
    \mathcal{B} = \frac{F_* + F_\textrm{blend} } {F_* }
\end{equation}

where $F_*$ is the flux from the target and and $F_\textrm{blend}$ is the sum of all other contaminating stars \citep{Kipping2010}.

Our transit model had 18 free parameters in total: (1) $R_P/R_*$, the radius ratio of the planet to the star, (2) $\rho_*$, the density of the star, (3) $b$, the impact parameter, (4) $q_{1, \textrm{\textit{Kepler}}}$, (5) $q_{2, \textrm{\textit{Kepler}}}$, (6) $q_{1, \textrm{TESS}}$, $\&$ (7) $q_{2, \textrm{TESS}}$, the limb darkening parameters for \textit{Kepler} and TESS, (8) $\mathcal{B}_\textrm{TESS}$, the TESS blend factor, and (9-18)  $\tau_1$, $\tau_2$, $\tau_4$, $\tau_5$, $\tau_6$, $\tau_7$, $\tau_8$, $\tau_9$, $\tau_{30}$, $\tau_{32}$, the ten times of transit minima. The parameters and priors used in the \texttt{MultiNest} transit model can be seen in Table \ref{table: transit_params}. The phase folded transit model fit to the ten transit observations from all five instruments can be seen in Figure \ref{fig: phase_fold_all}.

\begin{figure*}
    \centering 
    \includegraphics[width=\textwidth]{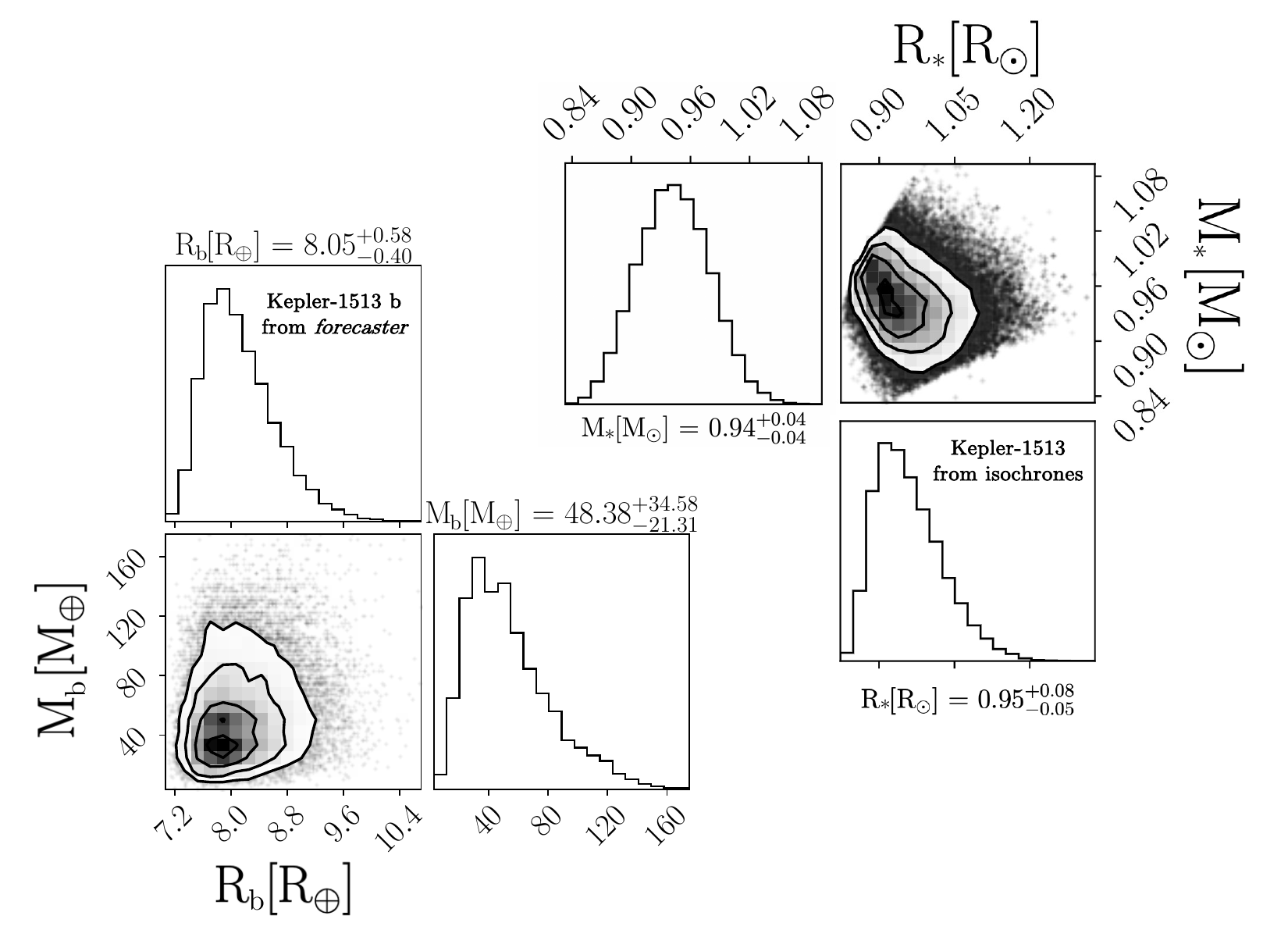}
    \caption{\textbf{Right:} Posterior results for the stellar mass and radius of Kepler-1513 from the \texttt{isochrone} package. \textbf{Left:} Posterior results for the planetary mass and radius of Kepler-1513\,b from the \texttt{forecaster} package. Both plots use the \texttt{corner} package \citep{corner}.}
    \label{fig: masses_and_radii}
\end{figure*}

\subsection{Determining the Physical Nature of Kepler-1513\hspace{0.15em}b}

Before we could further investigate the TTVs observed in the ten transits spanning more than a decade, we needed to determine the physical nature of Kepler-1513\,b, extrapolating from the transit model. 

First, we investigated the qualities of the host-star, Kepler-1513, using the Dartmouth Isochrones \citep{Dotter2008}.  We took the \textit{Gaia} DR3 parallax \citep{Luri2018}, the \textit{Kepler}
bandpass apparent magnitude, and the stellar atmosphere properties reported in the \textit{Kepler} DR25 catalog \citep{Mathur2017}, and appended them to a star.ini file along with their associated errors. These were then passed into the \texttt{isochrones} package \citep{Morton2015} to obtain \textit{a-posteriori} fundamental stellar parameters that are used later in our analysis for deriving planet/moon radii/masses. The mass and radius posteriors for Kepler-1513 can be seen in Figure \ref{fig: masses_and_radii}. The stellar parameters are reported in Table \ref{table: stellar_params}.

Next, we used \texttt{forecaster} to predict the mass of Kepler-1513\,b, based on the $R_P$/$R_*$ from a transit fit to the eight \textit{Kepler} transits \citep{ChenKipping2017}. We elected to use the $R_P$/$R_*$ value from the \textit{Kepler} transits only as the data is significantly more precise than the other LCs -- and it is more reliable to use a single homogenous dataset. The \texttt{forecaster} posterior results for the mass and radius of Kepler-1513\,b, excluding brown dwarf solutions, can also be seen in Figure \ref{fig: masses_and_radii}.

\begin{table}
\centering
\renewcommand{\arraystretch}{1.3}  

\caption{Stellar Parameters for Kepler-1513. Observational stellar parameters are from \textit{Gaia} DR3 (parallax) \citep{Luri2018} and \textit{Kepler} DR25 \citep{Mathur2017}. Fundamental stellar parameters are outputs from \texttt{isochrones} package \citep{Morton2015}, which uses the Dartmouth Isochrones \citep{Dotter2008} Credible interval derived from the [16$\%$, 50$\%$, \& 84$\%$] quantiles of the \textit{a-posteriori} solution.
}
\label{table: stellar_params}
\begin{tabular}{ c  c  c }
\hline
Parameter & Definition & Credible Interval\\
\hline
 & \textbf{Observational Stellar Parameters} & \\
$K_p$ & \textit{Kepler} Apparent Magnitude & 12.888 $\pm$ 0.100 \\
$T_\textrm{eff}$ [K] & Stellar Effective Temperature & 5491 $\pm$ 100  \\
$\log_{10}(g\,[\mathrm{cgs}])$ & Stellar Surface Gravity & 4.46 $\pm$ 0.10  \\
$\textrm{[M/H] [dex]}$ & Stellar Metallicity & 0.17 $\pm$ 0.06 \\
$\pi$ [mas] & Parallax  & 2.845 $\pm$ 0.013  \\

\hline
 & \textbf{Fundamental Stellar Parameters} & \\
$M_*$ [$M_\odot$] & Stellar Mass & $0.943^{+0.037}_{-0.037}$ \\
$R_*$ [$R_\odot$] & Stellar Radius & $0.950^{+0.077}_{-0.055}$  \\
$L_*$ [$L_\odot$] & Stellar Luminosity & $0.743^{+0.148}_{-0.100}$  \\
Age [Gyr] & Stellar Age & $7.0^{+4.0}_{-4.2}$  \\
\hline

\end{tabular}
\end{table}

\subsection{Lomb-Scargle Periodogram}

We fit a Lomb-Scargle (LS) periodogram \citep{VanderPlas2018} to Kepler-1513\,b's transit time minima from the transit model described above in order understand the periodicty of the TTVs. In order to recover both the short period and long period TTVs ($P_{TTV}$ and $\bar{P}_{TTV}$), we ran a Lomb-Scargle fit in two steps. 

First, we fit a Lomb-Scargle periodogram using linear regression with \texttt{numpy.linalg.solve} to solve the linear equation $F(x, \tau, P, \alpha_{TTV}, \beta_{TTV}, P_{TTV})$, over a TTV period grid $P_{TTV}$ with a range of 2-100 transits of Kepler-1513\,b evenly spaced in frequency space, where

\begin{equation} \label{eq: ls1}
\begin{aligned} 
    F(x, \tau, P, \alpha_{TTV}, \beta_{TTV}, P_{TTV}) = \\ \tau + P\,x + \alpha_{TTV}\sin \Big(\frac{2\pi x}{P_{TTV}}\Big)+ \beta_{TTV}\cos \Big(\frac{2\pi x}{P_{TTV}}\Big).
\end{aligned}
\end{equation}

Here, $x$ is the epoch number and P is the linear ephemeris transit period. This first Lomb-Scargle periodogram recovers the short period fast TTV ($P_{TTV}$), with a peak period of $\sim$2.6 epochs of Kepler-1513\,b, as the eight \textit{Kepler} transits, with higher precision transit minima dominate the signal. 

Second, we fit a second Lomb-Scargle periodogram, with two sinusoidal components over the same TTV period grid for one of the sinusoids ($\bar{P}_{TTV}$), but with the fast TTV period, $P_{TTV}$, as a free parameter initialized at its value corresponding with the maximum $\Delta \chi^2$ from the first single sinusoid Lomb-Scargle periodagram. We now must use non-linear least squares reduction, with \texttt{scipy.optimize.curve\_fit}, to solve for $F(x, \tau, P, \alpha_{TTV}, \beta_{TTV}, P_{TTV}, \bar{\alpha}_{TTV}, \bar{\beta}_{TTV}, \bar{P}_{TTV})$, as the equation is no longer linear, where

\begin{equation} \label{eq: ls2}
\begin{aligned}
    F(x, \tau, P, \alpha_{TTV}, \beta_{TTV}, P_{TTV}, \bar{\alpha}_{TTV}, \bar{\beta}_{TTV}, \bar{P}_{TTV}) = \\ \tau + P\,x + \alpha_{TTV}\sin \Big(\frac{2\pi x}{P_{TTV}}\Big)+ \beta_{TTV}\cos \Big(\frac{2\pi x}{P_{TTV}}\Big) + \\   \bar{\alpha}_{TTV}\sin \Big(\frac{2\pi x}{\bar{P}_{TTV}}\Big)+ \bar{\beta}_{TTV}\cos \Big(\frac{2\pi x}{\bar{P}_{TTV}}\Big).
\end{aligned}
\end{equation}

This second Lomb-Scargle periodogram reveals the new long period slow TTV, with an undetermined peak value that peaks and then flattens out due to lack of observational baseline around 10 cycles or epochs of Kepler-1513\,b. 

\begin{figure*}
\centering 

\includegraphics[width=.9\textwidth]{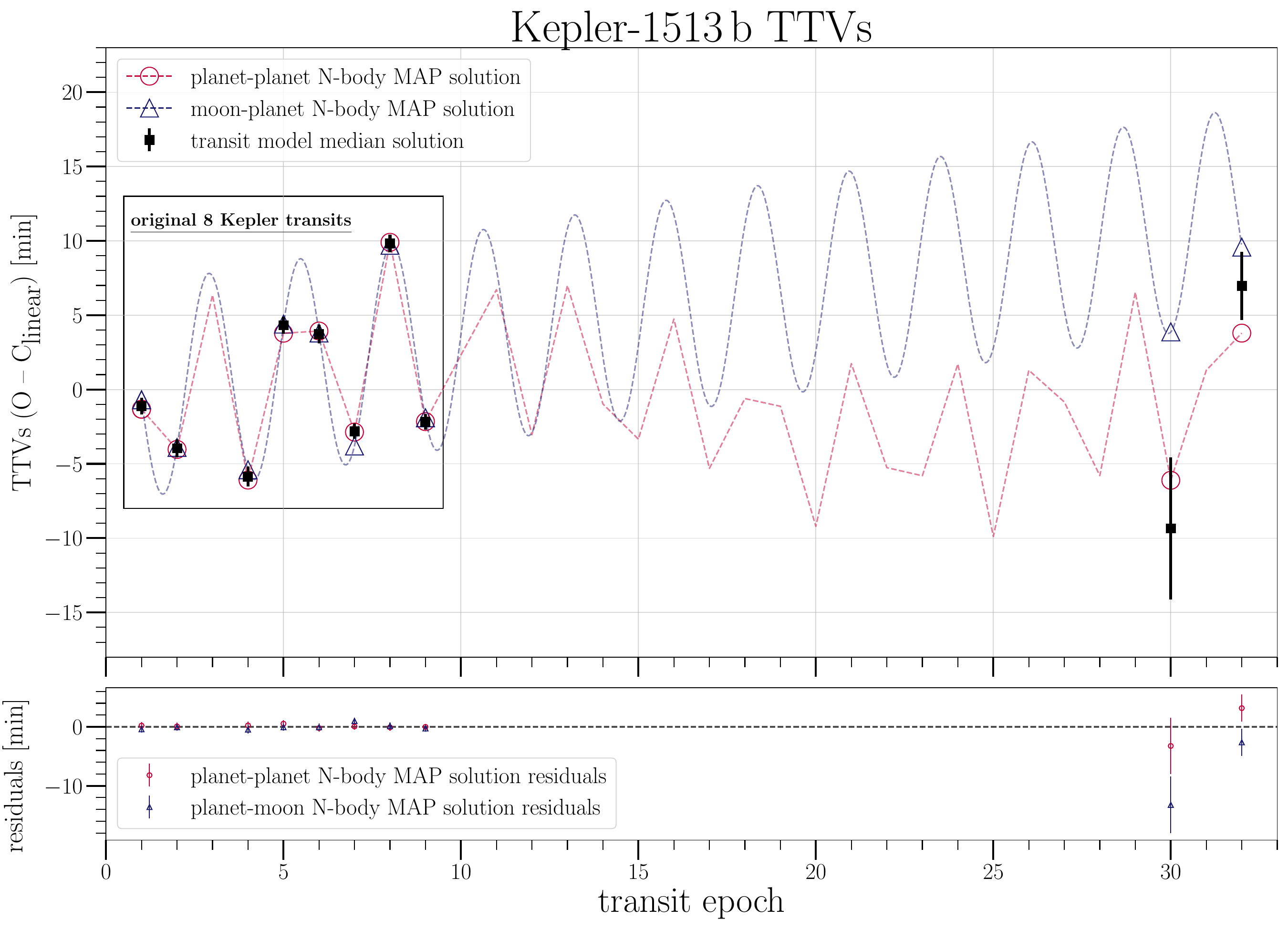}
\caption{Black data points with error bars are the TTVs determined from a transit model fit to the ten observed transits of Kepler-1513\,b (eight from Kepler, one from BARON + TESS, and one from LCOGT + Whitin). The red data points are the MAP solution from a planet-planet N-body fit to Kepler-1513\,b's transit times using \texttt{SWIFT} and \texttt{MultiNest}. The blue data points are the MAP solution from a planet-moon N-body fit to the Kepler-1513\,b's transit times using \texttt{LUNA} and \texttt{MultiNest}. The planet-moon fit fails to explain epoch 30.}
\label{fig: TTV_fit}

\includegraphics[width=.9\textwidth]{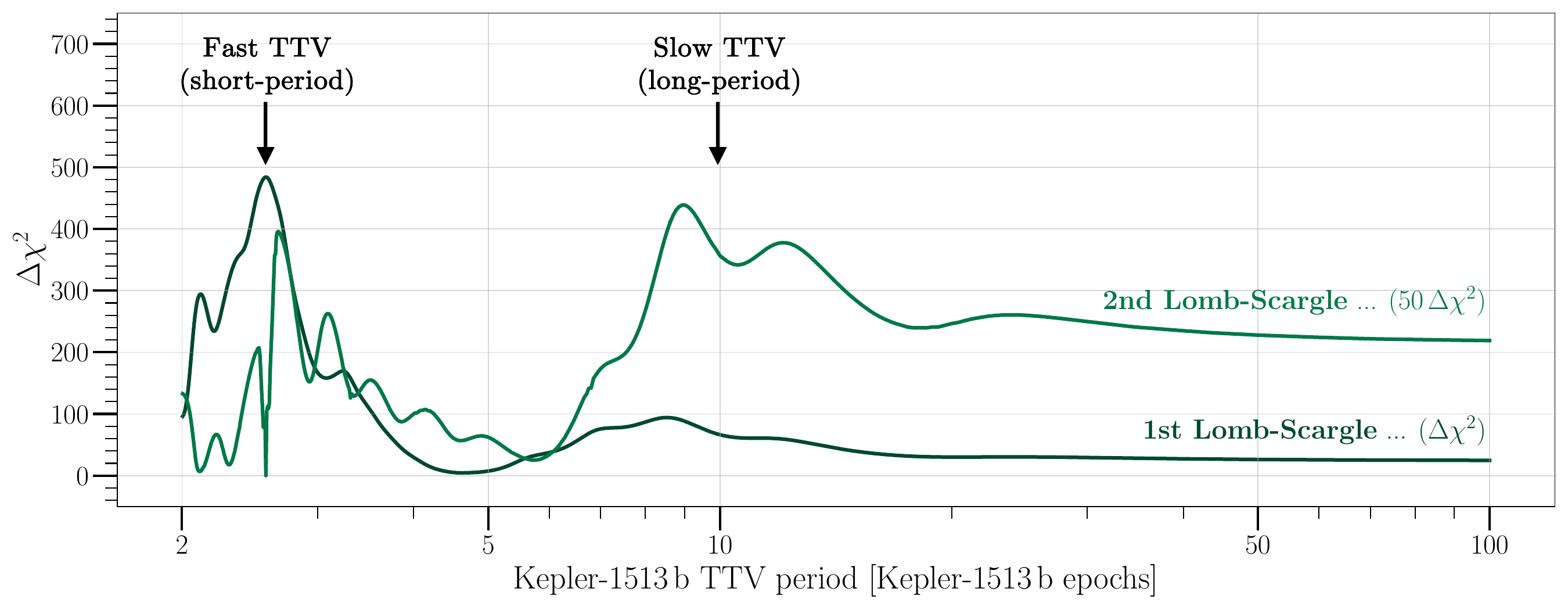}
\caption{Lomb-Scargle periodogram for all ten Kepler-1513\,b transits. Here we ran a Lomb-Scargle periodogram twice in order to recover both the fast TTV and the slow TTV signals. The first Lomb-Scargle returns the fast TTV signal, shown in the dashed line. We then fit a Lomb-Scargle periodogram with two sinusoidal waves, with one of the sinusoidal periods a free parameter, initialized at the fast period with the maximum $\chi^2$ value from this initial Lomb-Scargle periodogram. This second Lomb-Scargle periodogram, shown in the solid line, reveals the slow TTV signal. This shows that there are two periodic signals, (1) a fast TTV period that peaks around 2.6 epochs of Kepler-1513\,b and (2) a slow TTV signal that peaks and then flattens around 20-30 epochs of Kepler-1513\,b. The 2nd Lomb-Scargle (solid line that shows the slow TTV signal) $\Delta \chi^2$ values are all multiplied by 50, so the two curves can be viewed on the same axis. The functions minimized in our Lomb-Scargle periodogram can be seen in Equation (\ref{eq: ls1}) and Equation (\ref{eq: ls2}).}
\label{fig: lomb-scargle}

\end{figure*}

In order to determine the goodness of fit ($\Delta \chi^2$) values for our LS periodogram we also used linear regression to determine the optimal values assuming a linear ephemeris for  the transit times, our null model, to determine $\chi^2_{null}$. For each TTV period value in our grid, we determine the $\chi^2$ value for the TTV model ($\chi^2_{TTV}$) and define our goodness of fit ($\Delta \chi^2$) as $\Delta \chi^2$ = $\chi^2_{null}$ - $\chi^2_{TTV}$. Both Lomb-Scargle periodograms can be seen in Figure \ref{fig: lomb-scargle} -- note that here the $\Delta \chi^2$ values for the second Lomb-Scargle curve have been multiplied by fifty so that both curves would have a similar scale for plotting purposes. This analysis recovers two statistically significant independent periods in our full transit data.

\subsection{Planet-Planet TTVs}

In order to determine the physical and orbital parameters of an unseen planetary companion (Kepler-1513\,c) that could cause the observed Kepler-1513\,b TTVs, we used the N-body simulator \texttt{SWIFT} \citep{Levison1994, Nesvorny2013}. As discussed, the TTV parameter space is typically highly multi-modal, so we explored the parameter space using nested sampling with \texttt{MultiNest}, which is an efficient and effective sampling method for multi-modal parameter spaces. 

We adopted broad priors for all orbital and physical parameters for the unseen planet as we have no prior expectations for the unseen world. Specifically, the priors and modelling parameters in our planet-planet N-body model with \texttt{SWIFT} and \texttt{MultiNest} can be seen in Table \ref{table: swift_params}.

\begin{table*}
\centering
\renewcommand{\arraystretch}{1.3}  

\caption{Parameters for the Kepler-1513 planet-planet N-body model using N-body simulator \texttt{SWIFT} \citep{Levison1994, Nesvorny2013} with \texttt{MultiNest} \citep{Feroz2009} exploration of the parameter space. Credible interval derived from the [16$\%$, 50$\%$, \& 84$\%$] quantiles of the \textit{a-posteriori} solution.}
\label{table: swift_params}
\begin{tabular}{ c  c  c  c }
\hline
Parameter & Definition & Prior & Credible Interval \\
\hline 
\multicolumn{4}{c}{\centering \textbf{Nested Sampling Model Parameters}} \\

log$_{10}(M_b / M_*)$ & log$_{10}$ of Ratio-of-Masses for Kepler-1513\,b & $\mathcal{N}$(-3.81, 0.24) & $-3.81^{+0.23}_{-0.22}$\\

$M_c / M_*$ & Ratio-of-Masses for Kepler-1513\,c & $\log_e\mathcal{U}$(10$^{-6}$, 10$^{-2}$) & $0.000270^{+0.000100}_{-0.000064}$ \\

$T_\textrm{ref}$ [KBJD] & Reference Time & $\mathcal{U}$(277, 278) & $277.50591^{+0.00035}_{-0.00033}$ \\

$\lambda_c$  [deg] & Mean Longitude of Kepler-1513\,c & $\mathcal{U}$(0, 360) &  $226.0^{+20.0}_{-11.0}$ \\

P$_b$ [days] & Period of Kepler-1513\,b & $\mathcal{N}$(160.884, 0.1) & $160.8842^{+0.0011}_{-0.0028}$ \\

P$_c$ [days] & Period of Kepler-1513\,c & $\log_e\mathcal{U}$(30, 3000) & $841.4^{+8.1}_{-5.3}$ \\

e$_b$ & Eccentricity of Kepler-1513\,b & $\mathcal{U}$(0, 0.9) & $0.306^{+0.093}_{-0.097}$\\

e$_c$ & Eccentricity of Kepler-1513\,c & $\mathcal{U}$(0, 0.9) & $0.125^{+0.018}_{-0.019}$\\

$\varpi_b$ [deg] & Longitude of Periapses of Kepler-1513\,b & $\mathcal{U}$(0, 360)  & $259^{+21}_{-13}$ \\

$\varpi_c$ [deg] & Longitude of Periapses of Kepler-1513\,c & $\mathcal{U}$(0, 360)  & $313^{+20}_{-13}$ \\

b$_c$ [R$_*$] & Impact Paramater of Kepler-1513\,c & $\mathcal{U}$(1, 200) &  $72^{+58}_{-49}$ \\

$\Omega_c$ - $\Omega_b$ [deg] & Difference Between Longitudes of the Ascending Nodes & $\mathcal{U}$(0, 360) & $20^{+32}_{-46}$ \\

\hline
\multicolumn{4}{c}{\centering \textbf{Inferred Parameters}} \\

$M_b$ [$M_\textrm{Jup}$] & Mass of Kepler-1513\,b & ... & $0.152^{+0.104}_{-0.061}$ \\

$M_c$ [$M_\textrm{Jup}$] & Mass of Kepler-1513\,c & ... & $0.266^{+0.098}_{-0.063}$ \\

$K_b$  [m s$^{-1}$] & Radial Velocity Semi-Amplitude of Kepler-1513\,b & ... & $6.3^{+4.3}_{-2.5}$ \\

$K_c$ [m s$^{-1}$] & Radial Velocity Semi-Amplitude of Kepler-1513\,c & ... & $5.8^{+2.2}_{-1.4}$  \\

$\alpha_b$ [$\mu$as] & Astrometric Signal of Kepler-1513\,b & ... &  $0.26^{+0.18}_{-0.11}$ \\

$\alpha_c$ [$\mu$as] & Astrometric Signal of Kepler-1513\,c & ... &  $1.32^{+0.48}_{-0.31}$  \\

\hline
\end{tabular}
\end{table*}

On top of the standard \texttt{SWIFT} N-body model, we built in stability constraints for multi-planet systems based on the Hill stability criteria \citep{Petit2018} (specifically only keeping systems where C$_\textrm{sys}$ $<$ C$_\textrm{crit}$ is satisfied). We also exclude chaotic multi-planet systems based on constraints discussed in \citet{Tamayo2021} and originally presented in \citet{HaddenLithwick2018} (specifically only keeping systems where Z$_\textrm{sys}$ $<$ Z$_\textrm{crit}$ is satisfied).

Additionally, we built in a likelihood penalty into our \texttt{SWIFT} model based on constraints from the photoeccentric effect \citep{Dawson2012, Kipping2012a}. In short, the photoeccentric effect is an eccentricity driven astrodensity profiling. We can define a variable, $\Psi$, such that:

\begin{equation}
    \Psi = \frac{\rho_\textrm{*, obs}}{\rho_\textrm{*, true}},
\end{equation}

where $\rho_\textrm{*, obs}$ is determined from the transit model and  $\rho_\textrm{*, true}$ is determined from the Dartmouth Isochrones. From the literature, $\Psi$ also equals:

\begin{equation}
    \Psi = \frac{(1 + e \sin \omega)^3}{(1 - e^2)^{3/2}}
\end{equation}

We can interpret this likelihood penalty as follows: $\Psi$ effectively describes the orbital speed of the planet as compared to that expected from a circular orbit. Anything discrepant from $\Psi$ = 1 means that the orbit is less consistent with a circular orbit -- with $\Psi$ $>$ 1 implying a faster orbit and $\Psi$ $<$ 1 implying a slower orbit. Therefore, we can add a standard likelihood penalty to our total likelihood function based on the value of $\Psi$ determined from the transit observations, that helps to constrain the eccentricity, e, and the argument of periapsis, $\omega$, of Kepler-1513\,b.  Again, we only use the \textit{Kepler} transits to determine our asymmetric Gaussian $\Psi$, as this is the most precise data. Our \textit{Kepler} determined $\log_e(\Psi)$ value is ($-0.32^{+0.31}_{-0.11}$). We implement a asymmetric log-Gaussian likelihood penalty on $\Psi$ in our planet-planet N-body model. 

Our N-body model has twelve free parameters: (1-2) $M_b / M_*$ and $M_c / M_*$, the mass ratios of the two planets, (3) $T_\textrm{ref}$, a reference time between a linear ephemeris expected first transit and the time of the first observed transit of Kepler-1513\,b, (4) $\lambda_c$, the mean longitude of Kepler-1513\,c, (5-6) $P_b$ and $P_c$, the periods of the two planets, (7-8) $e_b$ and $e_c$, the eccentricities of the two planets, (9-10) $\varpi_b$ and $\varpi_c$, the longitudes of the periapses of the two planets, (11) $b_c$, the impact parameter of the perturbing planet, and (12) $\Omega_c$ - $\Omega_b$, the difference between the two longitudes of the ascending nodes, ($\Omega_b$ is fixed at 270 deg). The twelve parameters and the priors adopted can be seen in Table \ref{table: swift_params}.

The impact parameter of planet, $b$, can be converted to the inclination, $i$, of an orbit following given the semi-major axis, of the orbit, $a$, via the following relationship: $ i = \cos^{-1}(b \frac{R_*}{a})$. $T_\textrm{ref}$ in this model can be interpreted as the time of transit minimum of the first transit ($\tau_1$), assuming a linear ephemeris -- or said differently if there were no TTVs in the system.

The fit converged with two modes, clustered around a near MMR external planet commensurable with a $\sim$5:1 period ratio. Both modes were $\sim$5$\%$ outside the 5:1 period ratio. In what follows, we adopt the 1st mode and present these median and 1$\sigma$ results.

It is important to note that we do not claim this mode is the unique TTV solution. Rather, this solution appears to maximize the posterior density and is fully compatible with the data presently in hand. The point of our work isn't to identify the landscape of all possible perturbing planet solutions, but rather to ask whether a perturbing planet can explain the data well (thus applying context for the exomoon model). We therefore emphasize that because we only modeled with wide priors on the unseen planet's period, and did not specifically investigate other targetted period commensurabilities, that while we present an orbital solution for Kepler-1513\,c, further analysis is needed to confirm that this is the optimal solution. Radial velocity observations could help to investigate this as well.

The first mode MAP solutions from the \texttt{SWIFT} planet-planet fit with \texttt{MultiNest} can be seen in Table \ref{table: swift_params}. We find that Kepler-1513\,b's TTVs are consistent with an external planet, Kepler-1513\,c, a $0.266^{+0.098}_{-0.063}$ $M_\textrm{Jup}$ planet on a $841.4^{+8.1}_{-5.3}$ day orbit. The posteriors of the \texttt{SWIFT} planet-planet model fit with \texttt{MultiNest} can be seen in Figure \ref{fig: swift_corner}. The posteriors of the first mode of the \texttt{SWIFT} planet-planet model fit with \texttt{MultiNest} can be seen in Figure \ref{fig: swift_corner_1mode}. In order to select this mode, In order to select this mode, we determined the MAP solution, and then removed posteriors from other modes.

\subsection{Planet-Moon TTVs}

\begin{figure*}
    \centering 
    \includegraphics[width=\textwidth]{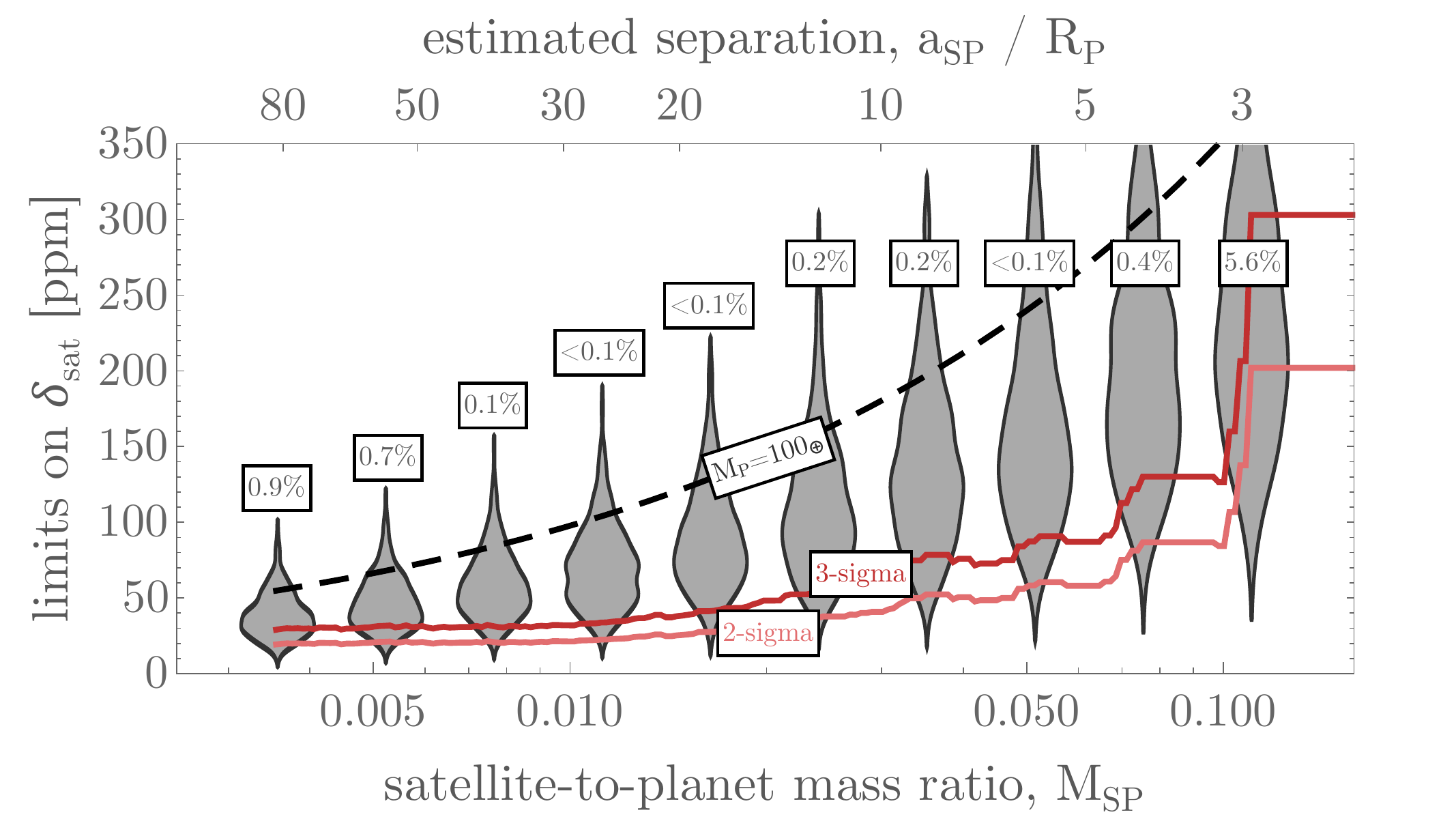}
    \caption{Transit origami performed on \textit{Kepler} photometry of Kepler-1513. Here we show the limits on the satellite (moon) transit depth vs. the satellite-to-planet mass ratio (bottom-axis) and the estimated separation (top-axis). The black dashed-line shows the expected transit depth for the satellite assuming a planet mass of $M_P = 100$\,$M_{\oplus}$, and then using \texttt{forecaster} to estimate the satellite radius, which can then be converted to transit depth.The violin plots show a full probabilistic \texttt{forecaster} model for the mass of Kepler-1513\,b based on the observed radius from the transit model. We were unable to detect any moon-like dips in the transit origami, and in red, we  plot the 2- and 3-$\sigma$ upper limit on the moon depth as a function of $M_S/M_P$ (and $a_{SP}/R_P$) as seen in the \textit{Kepler} photometry. Each violin plot also has the $p$-values representing the probability of the observations being consistent with a moon as a function of $M_S/M_P$ (and $a_{SP}/R_P$). We can see that the transit origami analysis, independent of the TTV N-body modeling, places significant tension on the moon hypothesis.}
    \label{fig:transit_origami}
\end{figure*}

\subsubsection{Photodynamical Model with \texttt{LUNA}} \label{section: LUNA}

We also tried to fit a planet-moon model to Kepler-1513\,b's transit times using the photodynamical code \texttt{LUNA} \citep{LUNA} and \texttt{MultiNest}. 

For this model, we have 17 free parameters. These are (1) $p$, the planet-to-star radius ratio ($R_P/R_*$), (2) $\rho_*$, (3) $b$, impact parameter for the planet, (4) $P_P$, the orbital period of the planet, (5) $\tau$, the time of first midtransit, (6) $q_{1, \textit{Kepler}}$, the first limb-darkening parameter for \textit{Kepler}, (7) $q_{2, \textit{Kepler}}$, the second limb-darkening parameter for \textit{Kepler}, (8) $q_{1, TESS}$, the first limb-darkening parameter for TESS, (9) $q_{2, TESS}$, the second limb-darkening parameter for TESS, (10) $\mathcal{B}_\textrm{TESS}$, the TESS blend factor, (11) $P_S$, the orbital period of the satellite, (12) $a_{SP}/R_P$, the satellite-to-planet semi-major axis divided by the planet radius, (13) $\phi_{s}$, the orbital phase of the satellite at the instant of planet–star inferior conjunction during the reference epoch, (14) $\cos{i_S}$, cosine of the satellite inclination, (15) $\Omega_S$ satellite longitude of ascending node, (16) $M_S/M_P$, the satellite-to-planet mass ratio, and (17) $R_S/R_P$ the satellite-to-planet radius ratio.

Uniform priors were adopted for all 17 free parameters, with three exceptions: $\rho_*$, $P_S$, and $\mathcal{B}_\textrm{TESS}$ were all sampled with log-uniform priors. We used a log-uniform prior on $\rho_*$ between $10^{-3}\,g\,cm^{-3}$ and $10^{3}\,g\,cm^{-3}$. For $P_S$, we adopted a log-uniform prior between 75 minutes and the period corresponding to one Hill radius. For $\mathcal{B}_\textrm{TESS}$, we use a log-uniform prior between 1 and 10. The semi-major axis of the satellite has a uniform prior from 2 to 100 planetary radii.

Despite repeated efforts to fit the data, \texttt{MultiNest} was not able to terminate under typical conditions of $>10^4$ posterior samples and instead $\sim$30,000 samples were returned. This is likely due to the model’s inability to explain the ninth transit observed by TESS + BARON (epoch 30 in Figure \ref{fig: TTV_fit}) and the poor likelihoods being found. Specifically, the MAP solution from the planet-moon model is inconsistent with the observed transit time of epoch 30 at nearly 3$\sigma$. This is evident from the planet-moon MAP TTV solution presented in Figure \ref{fig: TTV_fit}. Further, this is not surprising as a single perturbing moon is not expected to induce multiple TTVs periods in its host planet.

The LUNA posterior is quite multi-modal, as can be seen in the corner plot \ref{fig: luna_corner}. We extracted the MAP solution from the LUNA model and plot the produced TTVs in Figure \ref{fig: TTV_fit} vs. the TTVs from the transit model and the TTVs from the SWIFT model. As exomoon TTVs are expected to be sinusoidal, we also fit a sinusoid to the LUNA best fit times, which couldn't produce the two periods detectable in the transit model TTVs. This can be seen in Figure \ref{fig: lomb-scargle}. The posteriors for the best mode, surrounding the MAP solution in the \texttt{LUNA} model, are shown in Figure \ref{fig: luna_corner_1mode}.

\subsubsection{Transit Origami}

We also investigated the plausibility of a planet-moon model via the transit origami method \citep{Kipping2021_origami}. In modeling exoplanet transit photometry, phase-folding over the orbital period is often utilized to stack repeated transits of the planet. Unfortunately, this simple folding technique washes out the exomoon signal due to the moon's constantly changing phase \citep{Heller2019}. Transit origami solves for this problem by performing a kind of double-fold that correctly reconstructs a phase-coherent exomoon dip.

For this double fold, we restrict our analysis to the \textit{Kepler} photometry, since it is the only data for which a small moon transit could be plausibly recovered. Exomoon TTVs are expected to be sinusoidal \citep{Kipping2009_ttv_moon_I} with a period given by the aliased moon period \citep{Kipping2021} and this sets the second folding frequency for the transit origami technique. The phase of the planet's TTV (which we observe) is always opposite that of the moon's expected phase, and thus the position of the putative moon transit is calculable, modulo the unknown mass-ratio between the planet and the satellite, $(M_S/M_P)$. We thus search along a one-dimensional grid of mass ratios, for which at each position we know the TTV amplitude (from our TTV analysis), the mass ratio and thus also the corresponding planet-moon semi-major axis, $a_{SP}$ (since moon-induced TTVs $\propto (M_S/M_P) a_{SP}$).

We search along a 100-point log-uniform grid of mass ratios, at each step calculating the moon's expected mid-transit time, and then extracting the moon transit data assuming the transit duration equals that of the planetary transit. Only photometric points falling outside of the planetary transit are considered. We stack the final set of points occuring within the expected moon transit and measure its signal-to-noise ratio ala BLS \citep{Kovacs2002}. No significant dips are recovered, but we use the results to determine a 2- and 3-$\sigma$ upper limit on the moon depth as a function of $M_S/M_P$ (and $a_{SP}/R_P$) shown by the red lines in Figure~\ref{fig:transit_origami}.

It is instructive to compare these limits to that expected via mass-radius relations. Specifically, let assume the planet has a mass 100\,$M_{\oplus}$. In this case, for any given $(M_S/M_P)$ grid position we can calculate the corresponding $M_S$ value. We then feed this into the deterministic version of \texttt{forecaster} \citep{ChenKipping2017} for the terrestrial-worlds regime to compute a forecasted moon radius, and hence moon transit depth - shown by the black-dashed line in Figure~\ref{fig:transit_origami}. As one can see, this far exceeds the 3-$\sigma$ observational limit from transit origami, and thus we can exclude this hypothesis.

One might criticize the above procedure in that we somewhat arbitrarily assumed $M_P = 100$\,$M_{\oplus}$. Instead, let us now use \texttt{forecaster} in full probabilistic mode to predict a probability distribution for the mass of Kepler-1513\,b, then use these samples to calculate moon radii (again using probabilistic \texttt{forecaster}) and then finally make violin plots of the forecasted moon depth probability distribution, shown by the gray violins in Figure~\ref{fig:transit_origami}. Once again, it is visually apparent that the observed limits put strong pressure on these forecasts. Indeed, we label the $p$-values onto that figure above each violin.

The assumptions of transit origami are that the period of the moon is a few times greater than the transit duration (${\sim}$days) and the TTVs are solely induced by a single large moon. It should be noted that whilst indeed very close-in moons are not correctly modeled by the origami technique, they are somewhat irrelevant since they cannot produce large TTVs anyway. The expressions hold for general inclination/orientation positions of the moon's orbit, except for extreme cases where the moon's orbit is so wide and inclined it stops transiting. Against this hypothesis, our origami analysis places significant tension, and thus independently excludes the hypothesis of a single moon explaining the TTVs.

\subsubsection{More Complex Planet-Moon Models}
One could theoretically create a more complex solution involving multiple moons and likely explain that observed TTVs. However, following Occam's razor, we should prefer the solution with the lowest number of objects in the system, and thus such an analysis wasn't performed.

\subsection{Stellar Activity Induced TTVs}

\begin{figure}
    \centering 
    \includegraphics[width=.45\textwidth]{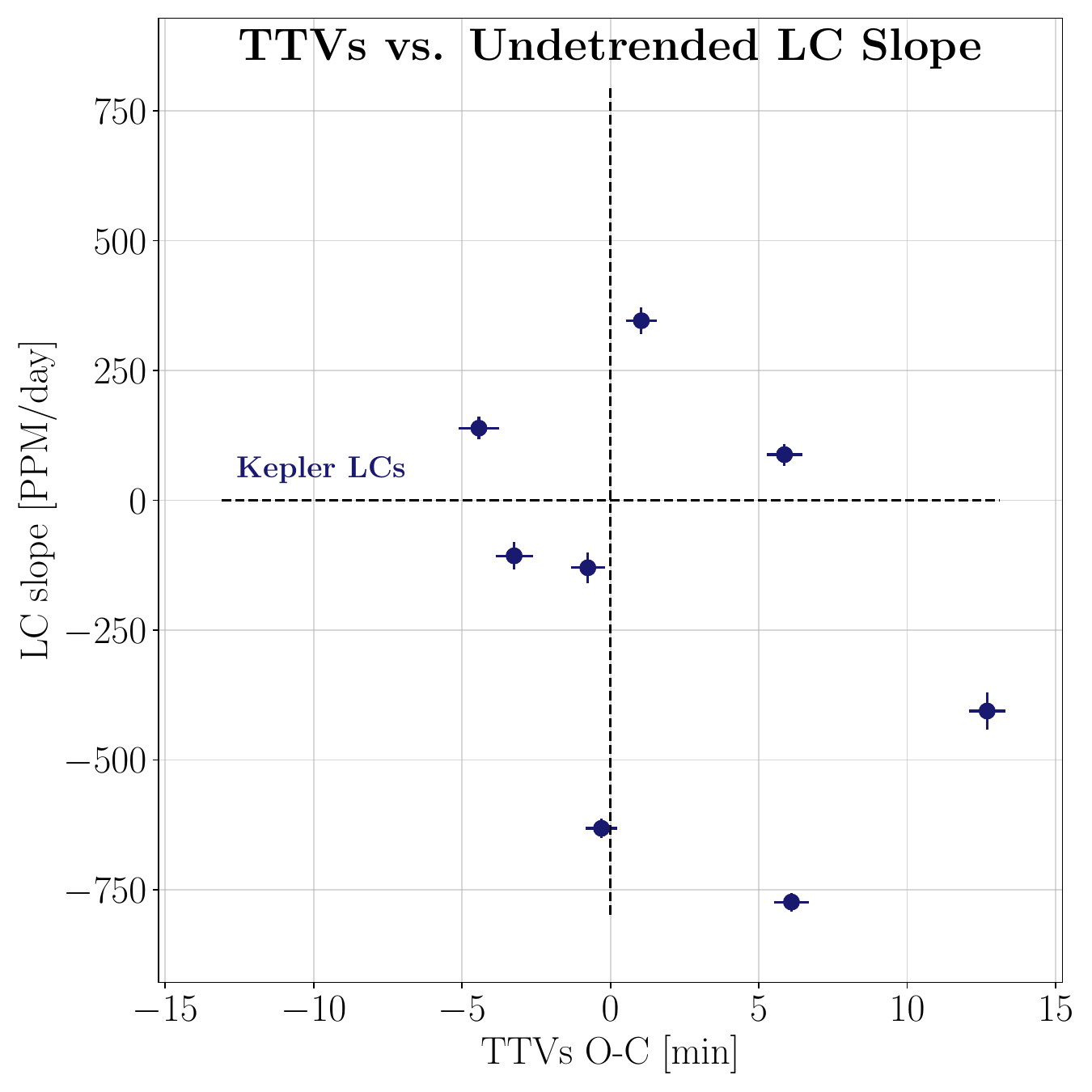}
    \caption{Light curve slope in parts per million (PPM) per day vs. TTVs in minutes. Slopes adopted from \citet{Holczer2015}. If the TTVs are induced by stellar activity, we expect a correlation between the TTVs and the LC slope. As we do not see that in this case we can be confident that the TTVs are caused by gravity of another world (planet or moon) in the planetary system rather than a star spot induced false positive.}
    \label{fig: starspot_ttvs}
\end{figure}

TTVs can be induced by stellar activity as well. Stellar activity induced TTV signals are, not-surprisingly, only expected around active stars. Kepler-1513 is a rotating variable star with a rotation period of $28.23 \pm 0.86$ days \citep{Mazeh2015b} and thus it is worth investigating whether the TTVs in this system could be induced by stellar activity. 

\citet{Mazeh2015} and \citet{Holczer2015} present a detailed explanation and analysis of stellar activity induced TTVs. In short, TTVs can be caused by stellar activity if a star spot crossing event occurs during ingress or egress of a transit. If the star spot crossing event happens at the beginning of the transit, then a late transit time will be observed as the spot will be moving from limb to center causing the start to get more faint. Therefore, if a late TTV occurs when the slope on either sides of the transit of the undetrended light curve is negative, then this could be a sign of a star spot induced TTV. Inversely, a star spot induced early TTV should coincide with a positive slope of the undetrending light curve on either sides of the transit. As a result, if the TTVs are induced by stellar activity one would expect two things: (1) the TTV period is equal to the stellar rotation period or an integer multiple of the stellar rotation period and (2) there would be a correlation between the slope of the undetrended LC and the TTVs.

First, we tested whether the observed TTV period is consistent with the observed stellar rotation period. The stellar rotation period 28.23 $\pm$ 0.86 days is less than the sampling period, which is equal to the transiting period of Kepler-1513\,b, or 160.88452 days. The TTVs will have a Nyquist period, or the minimum observable period, equal to twice the transiting period. Therefore, the stellar rotation period would be aliased if responsible for the TTVs. 

In order to determine the observable aliased period, we follow the same derivation as presented in \citet{McClellan1998}, and then adopted in \citet{Dawson2010} and subsequently in \citet{Kipping2021}. We find that the observed aliased TTV frequency peaks, $\nu$, in terms of the non-aliased physical TTV period, $P_{TTV}$, and the period of the transiting exoplanet, $P_\textrm{trans}$, occur at

\begin{equation}
    \nu = \Big|\frac{1}{P_{TTV}} \pm m \frac{1} {P_\textrm{trans}}\Big|
\end{equation}

where $m$ is a positive real integer and in this case $P_{TTV}$ equals $P_\textrm{rot}$ . Or, in terms of observed aliased TTV periods, P$'_{TTV}$, we have

\begin{equation} \label{eq: alias}
    P'_{TTV} = \frac{1}{\nu} = \frac{1}{\big|\frac{1}{P_{TTV}} \pm m \frac{1} {P_\textrm{trans}}\big|}
\end{equation}

Plugging into the Equation (\ref{eq: alias}), we find that only $m = 6$ with a negative sign returns a period larger than the Nyquist period, with a value of 534.61 days or 3.32 transit cycles. This aliased TTV period is not consistent with either the observed short period TTV signal ($\sim$2.6 cycles) or the long period TTV signal, which has a much longer period.

Next, we investigated whether there was any correlation between the TTVs and the LC slopes around transit. We adopted the slopes reported in \citet{Holczer2015} for Kepler-1513 in parts per million (PPM) per day and use the TTV values reported here. As shown in Figure \ref{fig: starspot_ttvs} there is no correlation between TTVs and the slope of the undetrended LCs -- therefore there is no reason to suspect that the TTVs are induced by stellar activity.

\section{Discussion}
\subsection{Understanding the Planet-Planet TTV Solution}

The MAP planet-planet TTV solution from \texttt{SWIFT} and \texttt{MultiNest} run with wide period priors was a $\sim$Saturn-mass planet, Kepler-1513\,c, slightly outside the 5:1 exterior near mean motion resonance (MMR) (period = 824.08) with Kepler-1513\,b (period = 160.88452 days). Assuming this solution, the long period TTV trend found in the data is caused by the super-period induced by the 5:1 near MMR, where the TTV super-period equation, as described in \citep{AgolFabrycky2018}, is

\begin{equation} \label{eq: super period}
    P_{TTV} = \frac{1}{|j/P_b - k/P_c|}.
\end{equation}

Here j and k are integers which represent the ratio of the MMR (so in this case, j is 1 and k is 5) and $P_b$ and $P_c$ are the two periods (so in this case $P_b$ is 160.8842 and $P_c$ is 841.4). Plugging into Equation (\ref{eq: super period}), we get an expected TTV period of $\sim$22.7 cycles of the transiter, which is consistent with the observations. 

Commonly, the short period ``chopping'' TTV effect is caused by the conjunctions of the two planets \citep{AgolFabrycky2018}. If chopping is caused by the conjunctions, one would expect the chopping to have a period equal to the synodic period, which equals

\begin{equation} \label{eq: conjunction period}
    P_\textrm{syn} = \frac{1}{|1/P_b - 1/P_c|}.
\end{equation}

Plugging in our MAP solution, our 5:1 MAP solution should cause a TTV period of 1.2 cycles of the transiter. Therefore, we would expect this conjunction period, which is less than the Nyquist period, to be aliased, if it is the cause of the short period TTVs. Using Equation (\ref{eq: alias}), we can test this hypothesis by recovering the aliased TTV period from the near 5:1 conjunction effect. Doing so, we find that only the choice $m = 1$ with a negative sign results in an aliased period larger than the Nyquist period. The aliased period, when adopting m = 1 with a negative sign, is $\sim$5.2 cycles, which is not consistent with the observed chopping, which has an observed period of $\sim$2.6 cycles. This is likely because in our 5:1 external near MMR, the two planets are so radially distant that the conjunction effect becomes negligible.

However, as explained in \citet{Nesvorny2009}, the TTVs induced by near MMR are explained in full by a summed linear equation of nearby resonances, with the amplitude of the TTV signal increasing as you approach resonance. In general, one can expect that the TTV super period to be dominated by a single near resonance term when within a few percent of a MMR \citep{Agol2005}, while Kepler-1513's MAP solution is $\sim$5$\%$ from 5:1 MMR. Using Equation (\ref{eq: super period}), we find the expected TTV super period from the 2:1 resonance of $\sim$1.6 transit cycles. This, as explained above, will be observed as an aliased period. Using Equation (\ref{eq: alias}), we again find that only m = 1 with a negative sign results in an aliased period larger than the sampling rate -- but, critically, we find an aliased TTV period of $\sim$2.6 transit cycles, which is consistent with the observed chopping effect. Therefore, the short period TTV is in fact explained by a second order chopping effect from the 2:1 near MMR aliased super period.

To summarize, we find that the observed TTVs can be explained by a $\sim$Saturn-mass planet $\sim$5$\%$ outside the 5:1 MMR. The first order TTV effect is explained by the 5:1 MMR super period of $\sim$22.7 transit cycles. The second order chopping effect is not caused by the planet conjunctions from the 5:1 distant orbit, but instead by an aliased super period from the distant 2:1 MMR.

\subsection{On the Future Confirmation of Kepler-1513\hspace{0.15em}c}

In obtaining two additional transits for Kepler-1513\,b we have discovered a previously unknown cause of planetary interlopers in the exomoon corridor: namely an insufficient observational baseline, for which the true long period TTV is undetectable. Further work should investigate how frequently this is expected to occur.

Additionally, while we have found a clean planet-planet TTV solution to Kepler-1513\,b's observed TTVs, the period space of planet-planet TTVs is multi-model. Similarly to the discovery of a second planet in the Kepler-19 system via TTVs, as presented in \citet{Ballard2011}, in this paper we identified a second planet in the system Kepler-1513, but we cannot decisively conclude and confirm the second planet's physical and orbital nature. Future work is needed to further investigate the planet-planet TTV period space and confirm the 5:1 external near MMR TTV solution. A closer analysis of other possible TTV solutions for Kepler-1513\,c would likely aid in disentangling the multi-modality present in TTV modeling

Perhaps, at least upper limits on secondary mass could be determined via radial velocity (RV) observations. As we have modeled the masses (ratios), periods, and eccentricities of both planets in the system, we can estimate the RV semi-amplitudes for both planets orbiting Kepler-1513, following

\begin{equation} \label{eq: rv_signal}
    K = \Big(\frac{2\pi G}{P}\Big)^{\frac{1}{3}} \frac{\textrm{M}_{P} \sin{i}}{(\textrm{M}_{P}+\textrm{M}_{*})^\frac{2}{3}} \frac{1}{\sqrt{1-e^2}},
\end{equation}

where i is the inclination, which equals

\begin{equation} \label{eq: inclination}
    i = \cos^{-1}\Big(b \frac{R_*}{a}\Big),
\end{equation}

and a is the semi-major axis, which can be derived using Kepler's third law. Using the posterior values from our Kepler-1513\,b transit model, presented in Table \ref{table: transit_params}, and the Kepler-1513 planet-planet TTV solution, presented in Table \ref{table: swift_params}, we estimate an RV semi-amplitude, for Kepler-1513\,b, $K_b$ = $6.3^{+4.3}_{-2.5} \, \textrm{m/s}$, and for Kepler-1513\,c, $K_c$ = of $5.8^{+2.2}_{-1.4} \, \textrm{m/s}$. Any future RV work should search for RV signals with these RV semi-amplitudes and the corresponding orbital periods. We also note that the stellar activity, as observed in the photometric data, may indeed challenge radial velocity observations, and thus also the subsequent ability to accurately and precisely model the RV signal.

Additionally, as both planets in our MAP planet-planet TTV solution appear to be cool gas giants, one might ask whether an astrometric detection may be possible for Kepler-1513\,b and  Kepler-1513\,c. Most obviously, one might consider detection via \textit{Gaia}'s full astrometry catalogue to be released in DR4, which will be released not before the end of 2025 \citep{Gaia2016}.\footnote{https://www.cosmos.esa.int/web/gaia} We can estimate whether \textit{Gaia} will be sensitive to Kepler-1513\,b and Kepler-1513\,c's astrometric signal by estimating $a_*$, the semi-major axis of Kepler-1513, the host star in the system, due to each planet. As presented in \citet{Perryman2014}, this can be determined by assuming a Keplerian orbit, via

\begin{equation}
    a_* = a \, \Big(\frac{M_P}{M_*}\Big).
\end{equation}

In turn, we can then convert $a_*$ to $\alpha$, the corresponding quantity in angular measure, based on the distance to the star, $d$. This quantity, $\alpha$, is generally referred to as the astrometric signature, given by

\begin{equation}
    \alpha = \Big(\frac{a_*}{1 \, \textrm{AU}}\Big)\Big(\frac{d}{1 \, \textrm{pc}}\Big)^{-1} \, \textrm{arcsec}.
\end{equation}

We assumed a distance to Kepler-1513 of 351.5 pc, taken from the \textit{Gaia} EDR3 parallax \citep{Gaia2020_edr3}. This gives us an estimated $\alpha_b = 0.26^{+0.18}_{-0.11} \, \mu \textrm{as}$ and $\alpha_c = 1.32^{+0.48}_{-0.31} \, \mu \textrm{as}$, respectively. This is below the expected precision of \textit{Gaia} astrometry, and so, unfortunately \textit{Gaia} astrometry is unlikely to aid in this pursuit.

\section{Conclusion}
We present our work, investigating the cause of the short period TTVs observed in the \textit{Kepler} photometric observations of Kepler-1513\,b. Two additional transits observed using TESS, BARON, LCOGT, and Whitin (about a decade after the last \textit{Kepler} transit) significantly alters the TTV signal, revealing a long TTV super period, previously undetectable because the observational baseline was significantly less than the TTV period. The original short-period TTV signal is still present, but now manifests as a chopping signal on top of the longer super period. Using \texttt{SWIFT} and \texttt{MultiNest} we demonstrate that Kepler-1513\,b's TTVs are consistent with an external planet, Kepler-1513\,c, a $0.266^{+0.098}_{-0.063}$ $M_\textrm{Jup}$ planet on a $841.4^{+8.1}_{-5.3}$ day orbit. We also explore the possiblity that the TTV signal could still be caused be an exomoon, using both photodynamical modeling with \texttt{LUNA} and the transit origami technique, but find that a single planet-moon model fails to reproduce Kepler-1513's LCs in both models. We also find no evidence that these TTVs could be stellar activity induced. 

Therefore, we argue that the unseen companion around Kepler-1513 is likely to be a $\sim$Saturn mass planet on a wide orbit near the 5:1 MMR with Kepler-1513\,b, Kepler-1513\,c, but emphasize the need to further investigate other planet-planet TTV solutions. 

Finally, we take Kepler-1513\,c as a new example of a planetary interloper in the exomoon corridor, namely a short observational baseline induced undetectable super period. We suggest that future work should further study the period space of planet-planet TTVs and specifically how frequently planet-planet TTVs will be observed in the exomoon corridor due to an insufficient observational baseline.


\section*{Acknowledgements}

We thank the anonymous referee for comments that greatly improved this manuscript.

D.A.Y. and D.K. acknowledge support from NASA Grant \#80NSSC21K0960.

D.A.Y. thanks the LSSTC Data Science Fellowship Program, which is funded by LSSTC, NSF Cybertraining Grant \#1829740, the Brinson Foundation, and the Moore Foundation; his participation in the program has benefited this work.

P.D. acknowledges support by a 51 Pegasi b Postdoctoral Fellowship from the Heising-Simons Foundation and by a National Science Foundation (NSF) Astronomy and Astrophysics Postdoctoral Fellowship under award AST-1903811.

J.E. acknowledges the support of BRIEF (Boyce Research Initiatives and Education Foundation)
for use of its Boyce-Astro Research Observatories, BARO and BARON. (www.boyce-astro.org).

This work made use of \texttt{tpfplotter} by J. Lillo-Box (publicly available in www.github.com/jlillo/tpfplotter), which also made use of the python packages \texttt{astropy}, \texttt{lightkurve}, \texttt{matplotlib} and \texttt{numpy}. This research made use of Lightkurve, a Python package for \textit{Kepler} and TESS data analysis \citep{Lightkurve2018}. 

Analysis was carried out in part on the NASA Supercomputer PLEIADES (Grant \#HEC-SMD-17- 1386), provided by the NASA High- End Computing (HEC) Program through the NASA Advanced Supercomputing (NAS) Division at Ames Research Center. This research has made use of the NASA Exoplanet Archive, which is operated by the California Institute of Technology, under contract with the National Aeronautics and Space Administration under the Exoplanet Exploration Program. This paper includes data collected by the \textit{Kepler} Mission. Funding for the \textit{Kepler} Mission is provided by the NASA Science Mission directorate. This paper includes data collected by the TESS mission. Funding for the TESS mission is provided by the NASA's Science Mission Directorate. This work has made use of data from the European Space Agency (ESA) mission Gaia (\href{https://www.cosmos.esa.int/gaia}{https://www.cosmos.esa.int/gaia}), processed by the Gaia Data Processing and Analysis Consortium (DPAC, \href{https://www.cosmos.esa.int/web/gaia/dpac/consortium}{https://www.cosmos.esa.int/web/gaia/dpac/consortium}). Funding for the DPAC has been provided by national
institutions, in particular the institutions participating in the Gaia Multilateral Agreement.

This work makes use of observations from the LCOGT network. Part of the LCOGT telescope time was granted by NOIRLab through the Mid-Scale Innovations Program (MSIP). MSIP is funded by NSF.

This research has made use of the Exoplanet Follow-up Observation Program (ExoFOP; DOI: 10.26134/ExoFOP5) website, which is operated by the California Institute of Technology, under contract with the National Aeronautics and Space Administration under the Exoplanet Exploration Program.

KAC acknowledges support from the TESS mission via subaward s3449 from MIT.

KKM acknowledges support from the New York Community Trust Fund for Astrophysical Research.

This work was supported by patreons to the Cool Worlds Lab, for which the authors thank
D. Smith, M. Sloan, C. Bottaccini, D. Daughaday, A. Jones, S. Brownlee, N. Kildal, Z. Star, E. West, T. Zajonc, C. Wolfred, L. Skov, G. Benson, A. De Vaal, M. Elliott, B. Daniluk, M. Forbes, S. Vystoropskyi, S. Lee, Z. Danielson, C. Fitzgerald, C. Souter, M. Gillette, T. Jeffcoat, J. Rockett, D. Murphree, S. Hannum, T. Donkin, K. Myers, A. Schoen, K. Dabrowski, J. Black, R. Ramezankhani, J. Armstrong, K. Weber, S. Marks, L. Robinson, S. Roulier, B. Smith, G. Canterbury, J. Cassese, J. Kruger, S. Way, P. Finch, S. Applegate, L. Watson, E. Zahnle, N. Gebben, J. Bergman, E. Dessoi, J. Alexander, C. Macdonald, M. Hedlund, P. Kaup, C. Hays, W. Evans, D. Bansal, J. Curtin, J. Sturm, RAND Corp., M. Donovan, N. Corwin, M. Mangione, K. Howard, L. Deacon, G. Metts, G. Genova, R. Provost, B. Sigurjonsson, G. Fullwood, B. Walford, J. Boyd, N. De Haan, J. Gillmer, R. Williams, E. Garland, A. Leishman, A. Phan Le, R. Lovely, M. Spoto, A. Steele, M. Varenka, K. Yarbrough, A. Cornejo, D. Compos, F. Demopoulos, G. Bylinsky, J. Werner, B. Pearson, S. Thayer, T. Edris \& M. Waters.

We gratefully acknowledge the open source
software which made this work possible:
\texttt{AstroImageJ} \citep{Collins2017},
\texttt{celerite2} \citep{celerite1, celerite2},
\texttt{corner} \citep{Foreman-Mackey2016}, 
\texttt{exoplanet} \citep{Foreman-Mackey2021}, 
\texttt{matplotlib} \citep{matplotlib},
\texttt{numpy} \citep{numpy}, 
\texttt{PyMC3} \citep{Salvatier2016},
\texttt{scipy} \citep{scipy},
\texttt{SWIFT} \citep{Levison1994, Nesvorny2013},
\texttt{TAPIR} \citep{Jensen2013}.

\section*{Data Availability}

The code used and results generated by this work are made publicly available at \href{https://github.com/dyahalomi/Kepler1513}{https://github.com/dyahalomi/Kepler1513}.



\bibliographystyle{mnras}
\bibliography{main} 



\appendix
\section{\texttt{democratic\_detrender}} \label{sec: democratic_detrender}
\begin{figure*}
\centering 

\includegraphics[width=.95\textwidth]{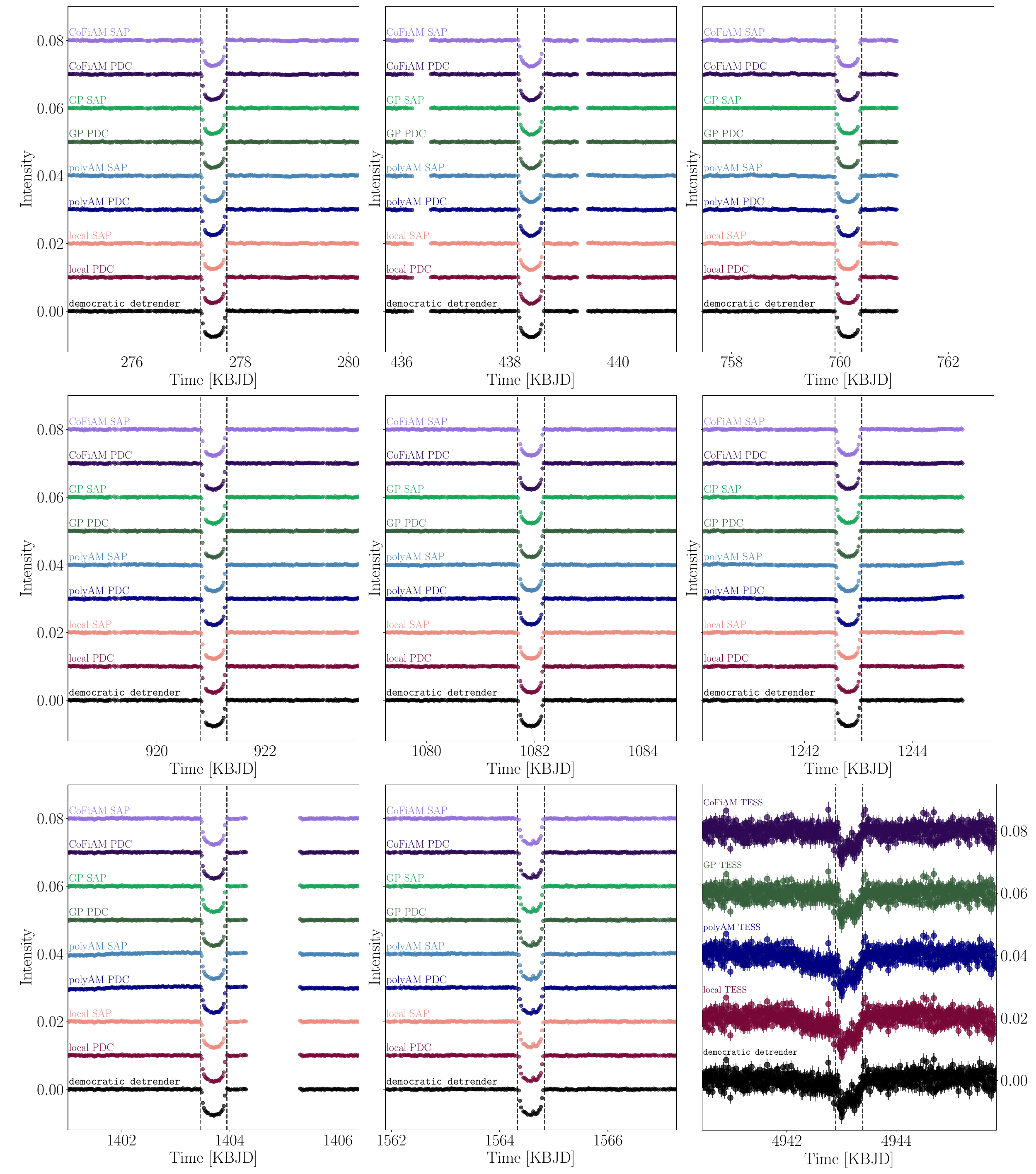}
\caption{Democratically detrended \textit{Kepler} and TESS light curves for Kepler-1513\,b all in KBJD = BJD - 2,454,833 days. In each subplot, each color represents a specific detrending method ran on a specific photometric dataset for a given transit epoch, and the the bottom transit is the final democratically detrended (or method marginalized detrended) transit observation used for modeling. For each photometric dataset, we detrend using four detrending methods \texttt{CoFiAM}, \texttt{polyAM}, \texttt{local}, and \texttt{GP}. For the \textit{Kepler} data, we  detrended both the ``Pre-search Data Conditioning (PDC) light curves and the ``Simple Aperture Photometry'' (SAP) light curves and then take the median value of all eight LCs at each time step to determine the democractically detrended value.}

\label{fig: detrended_lc}

\end{figure*}

Figure \ref{fig: detrended_lc} shows a sample output using the \texttt{democratic\_detrender} to detrend the \textit{Kepler} and TESS photometry of Kepler-1513\,b. The \texttt{democratic\_detrender} uses four different detrending methods in detrending stellar photometry:

\begin{itemize}
    \item \texttt{CoFiAM} or Cosine Filtering with Autocorrelation Minimisation builds on cosine filtering approach used to study CoRoT data \citep{Mazeh2010}. In \texttt{CoFiAM}, we train 30 models with N cosines in our fit, where N ranges from 1 to 30, and at each epoch pick the cosine filter that leads to the least correlated light curve via the Durbin-Watson statistic \citep{DurbinWatson1950, Kipping2013c}.
    
     \item \texttt{polyAM} or Polynomial detrending with Autocorrelation Minimisation follows a similar process to \texttt{CoFiAM}, except we train 30 models with polynomial models as the basis function. Polynomial filtering is a common method for stellar activity detrending \citep{Fabrycky2012, Gautier2012, Giles2018}. In \texttt{polyAM}, the 30 different bases models are 1$^\textrm{st}$- to 30$^\textrm{th}$-order polynomials. For each epoch, as above, the least correlated light curve via the Durbin-Watson statistic \citep{DurbinWatson1950, Kipping2013c} is chosen.
     
      \item \texttt{local} method again uses 1$^\textrm{st}$- to 30$^\textrm{th}$ order polynomials, but the \texttt{local} detrended light curve is selected via the lowest Bayesian Information Criterion \citep{Schwarz1978} computed on the data within six transit durations of the time of mid-transit. This a fairly typical detrending method for the analysis of short-period transiters \citep{Sandford2017}.
      
       \item \texttt{GP} or Gaussian process regression, as the name suggests, uses a Gaussian process to detrend the stellar activity. We used a quasiperiodic Gaussian process, as it has been shown that it is possible to model stellar activity of a rotating star using a quasiperiodic kernel \citep{Angus2017}. Specifically, we used a SHOTerm kernel from \texttt{celerite2} via the \texttt{exoplanet} package, which is a stochastically-driven, damped harmonic oscillator \citep[\texttt{exoplanet},][]{Foreman-Mackey2021}, \citep[\texttt{celerite2},][]{celerite1, celerite2}.
\end{itemize}

For the \textit{Kepler} data, we apply these four methods to both the ``Pre-search Data Conditioning (PDC) light curves and the ``Simple Aperture Photometry'' (SAP) light curves. For the TESS data, we apply these four methods to the light curves that we extracted from the photometry. Before each detrending method is applied to the data, we mask the transits and remove all data points that are more than 4-$\sigma$ deviants from a moving median of bandwidth 30 cadences outside of transit. We then detrend the masked LCs using all four methods and extrapolate for the transits. Finally, we adopt the median of each time series data point as the democratically detrended LC value. Additionally, we inflate the uncertainty on each photometric data point by adding in quadrature the reported errors with 1.4286 multiplied by the median absolute deviation (MAD) between the individual detrending models. This 1.4286 value is the constant scale factor to convert from MAD to standard deviation assuming Gaussian errors. This allows us to pass information into the detrended LC about how well the different detrending methods agreed at each data point. MAD is used in the place of standard deviation to mitigate the influence of possible failed detrending(s).

This democratic detrending package is accessible via GitHub.\footnote{\href{https://github.com/dyahalomi/democratic\_detrender}{https://github.com/dyahalomi/democratic\_detrender}}

\section{\texttt{SWIFT} + \texttt{MultiNest} Corner Plots}
\begin{figure*}
\centering 

\includegraphics[width=\textwidth]{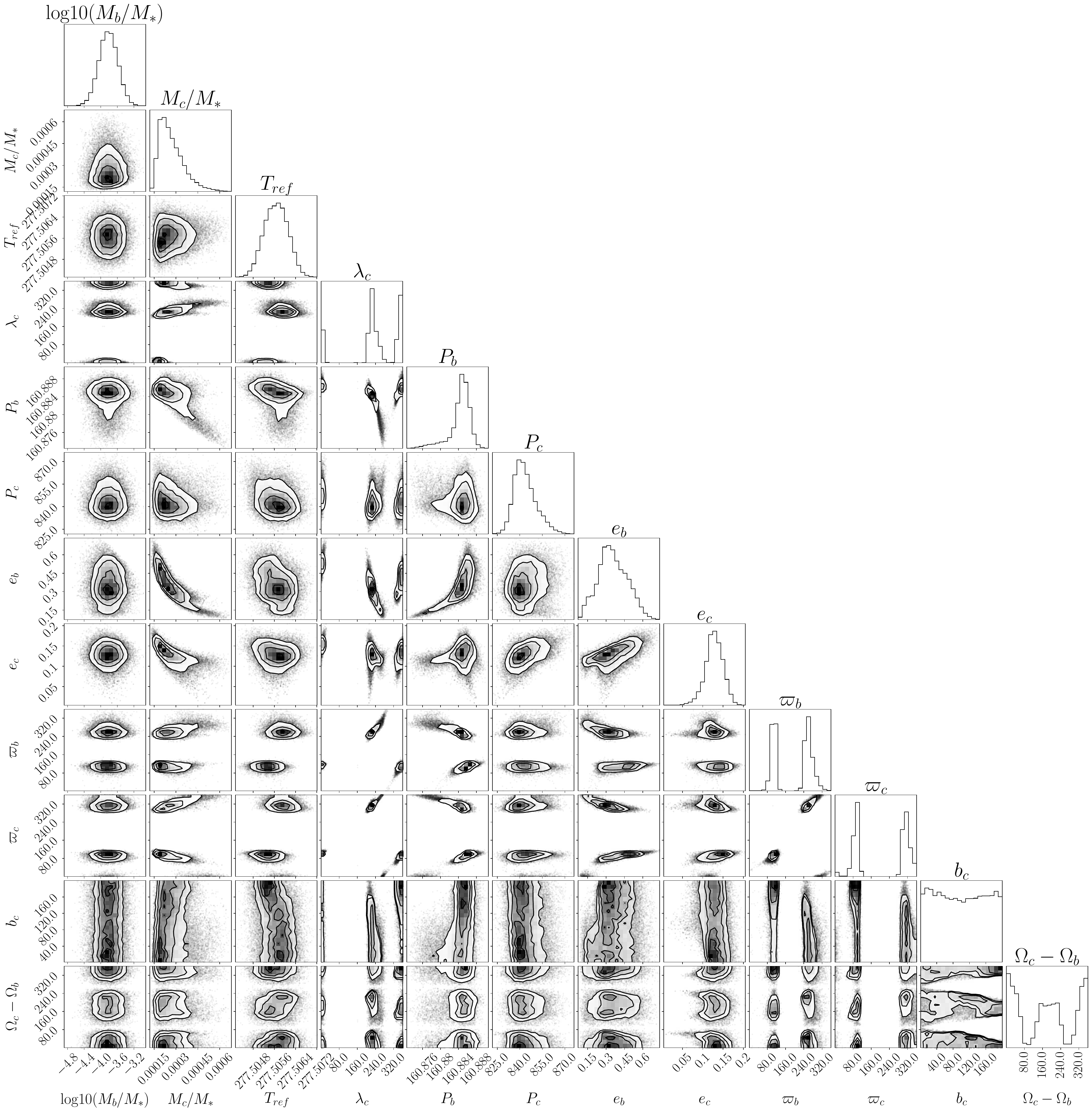}
\caption{Corner figure for the \texttt{SWIFT} planet-planet model posteriors, sampled using \texttt{MultiNest}, and with priors presented in Table \ref{table: swift_params}. The multi-modality of planet-planet TTVs is clearly visible in the posteriors.}
\label{fig: swift_corner}

\end{figure*}

\begin{figure*}
\centering 

\includegraphics[width=\textwidth]{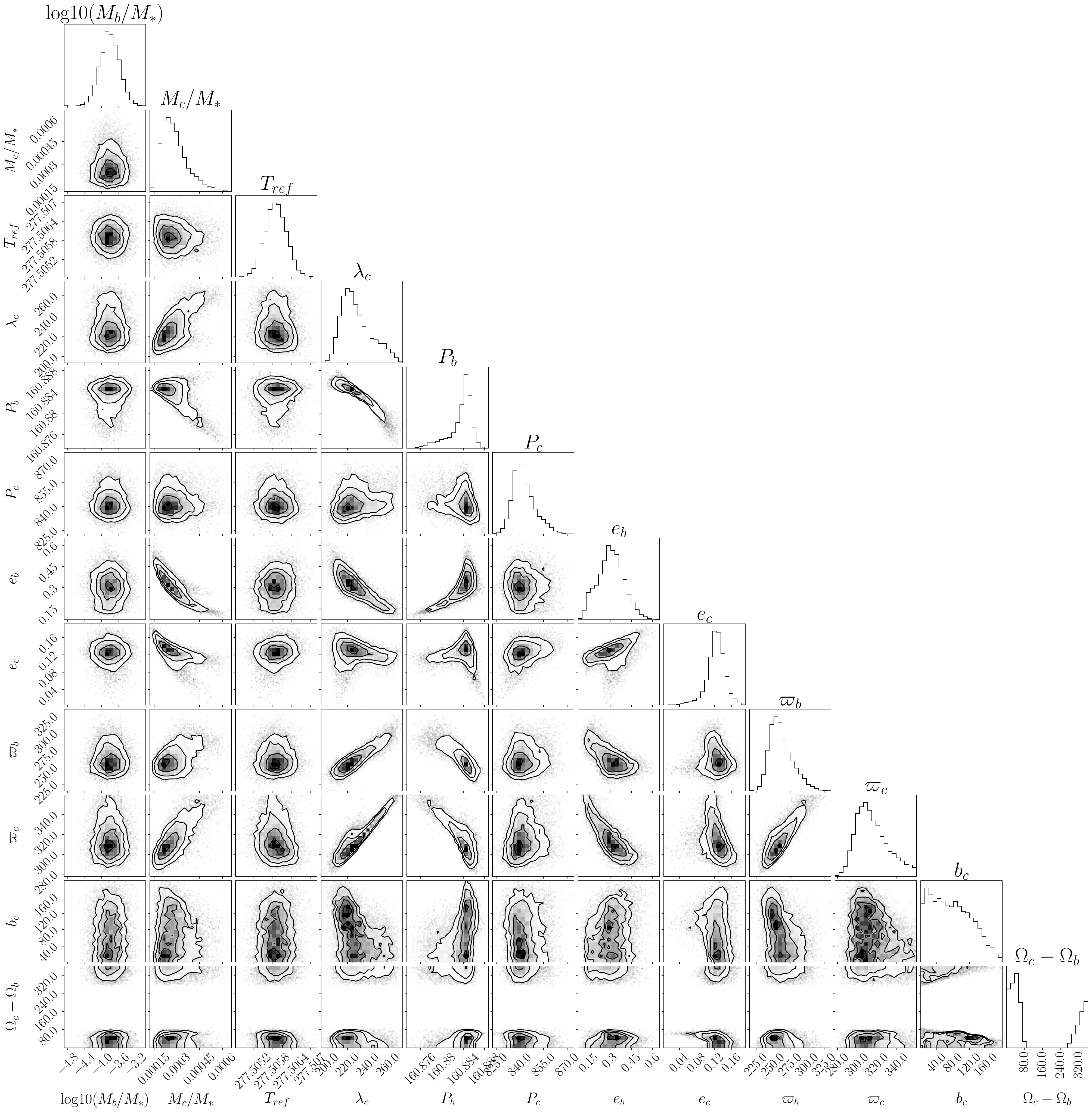}
\caption{Corner figure for the first mode of the \texttt{SWIFT} planet-planet model posteriors, sampled using \texttt{MultiNest}, and with priors presented in Table \ref{table: swift_params}. In order to select this mode, we determined the MAP solution, and then removed posteriors from other modes. Note the $\Omega_c - \Omega_b$ posterior is a single peak, wrapped around the sampling limits of the parameter (0 to 360 degrees).}
\label{fig: swift_corner_1mode}

\end{figure*}

Below are the corner plots showing the posterior chains from the \texttt{SWIFT} planet-planet N-body model sampled with \texttt{MultiNest}. Figure \ref{fig: swift_corner} shows the full posterior, while Figure \ref{fig: swift_corner_1mode} only shows the best-fit mode based on including the MAP solution.

\section{\texttt{LUNA} Planet-Moon Model} \label{sec: luna_model}
\begin{figure*}
\centering 

\includegraphics[width=\textwidth]{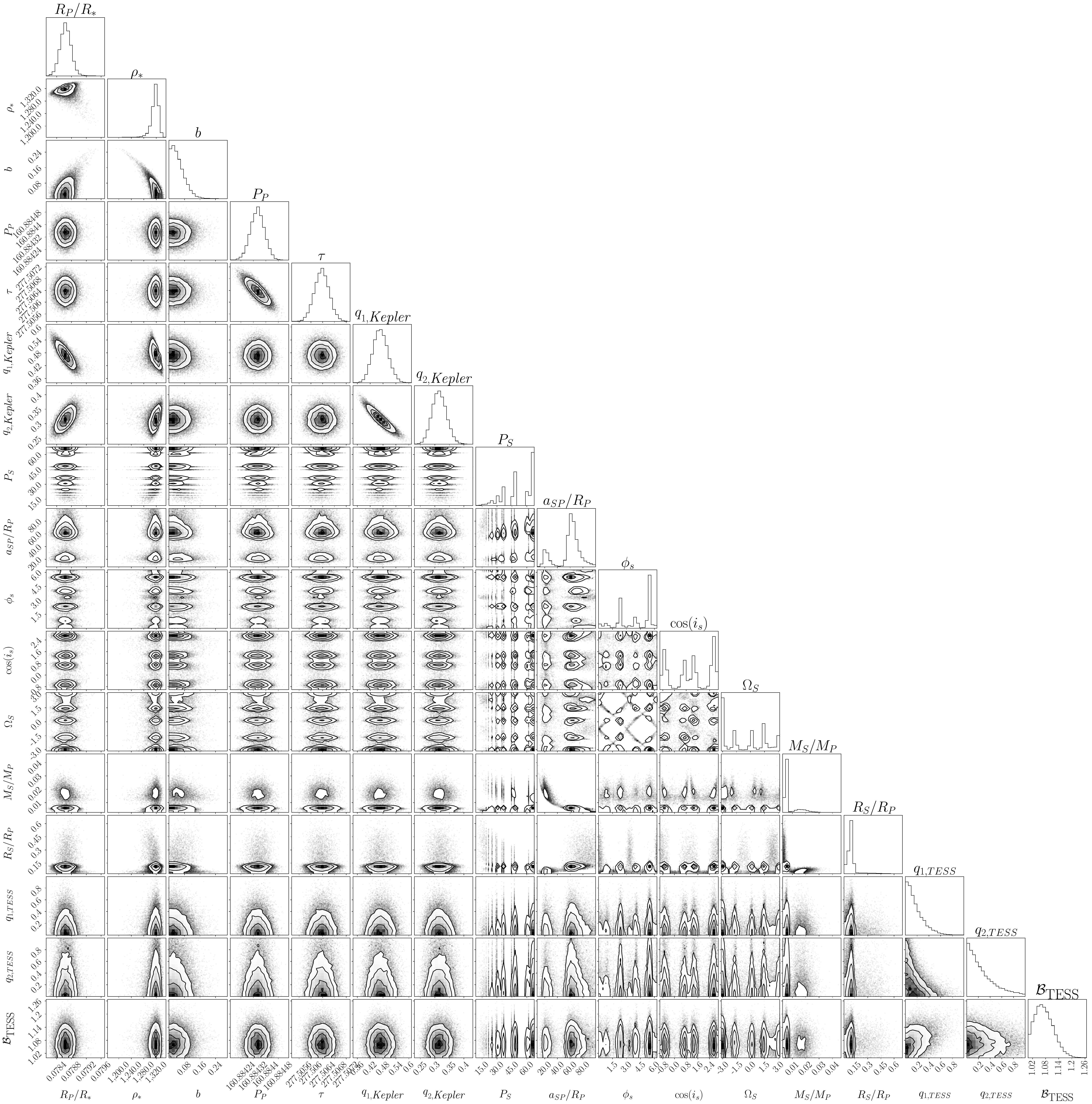}
\caption{Corner figure for the \texttt{LUNA} planet-moon model posteriors, sampled using \texttt{MultiNest}, and with priors presented in Section \ref{section: LUNA}. The multi-modality of planet-moon TTVs is clearly visible in the posteriors.}
\label{fig: luna_corner}

\end{figure*}

\begin{figure*}
\centering 

\includegraphics[width=\textwidth]{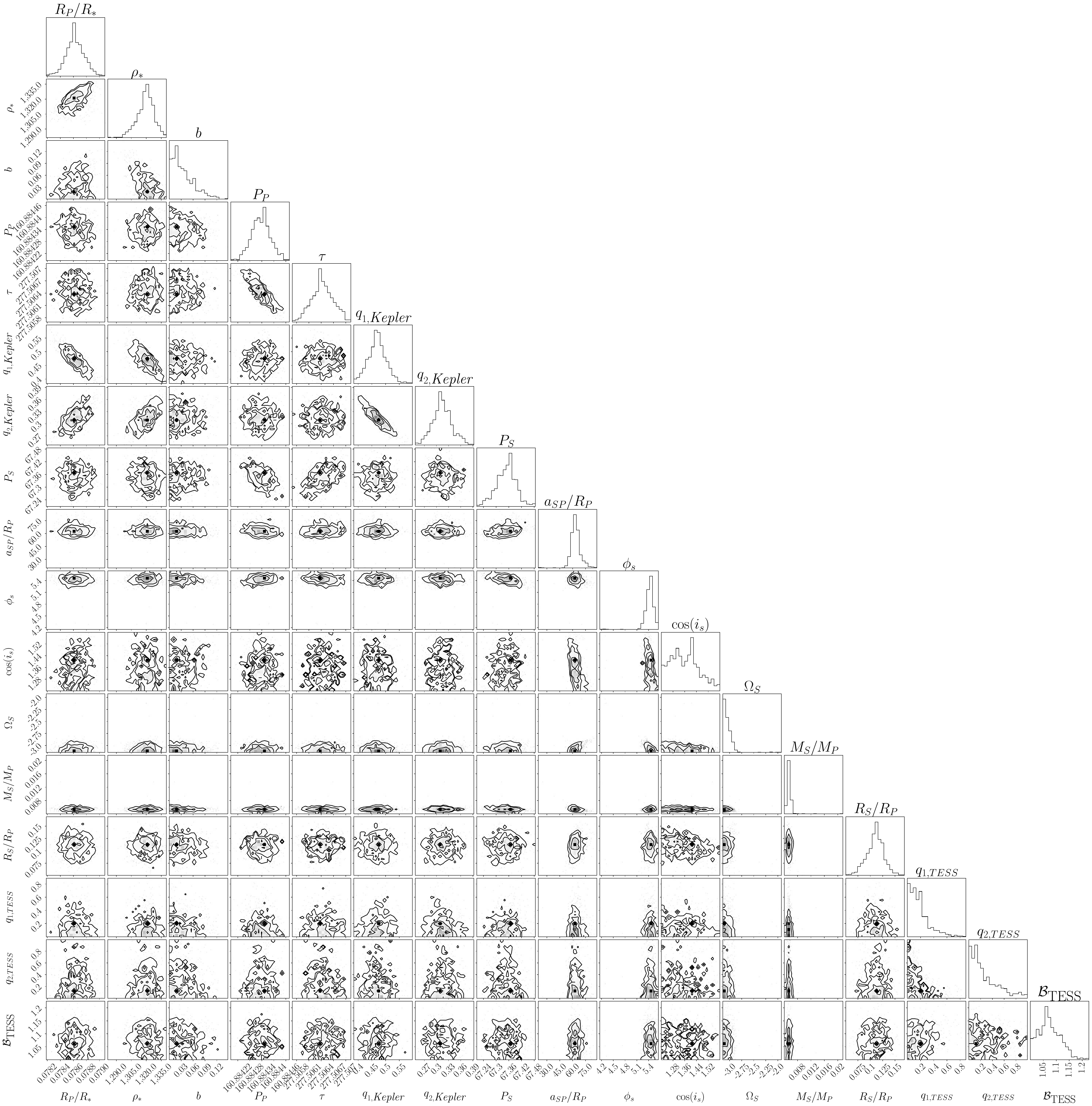}
\caption{Corner figure for the first mode of the \texttt{LUNA} planet-moon model posteriors, sampled using \texttt{MultiNest}, and with priors presented in Section \ref{section: LUNA}. In order to select this mode, we determined the MAP solution, and then removed posteriors from other modes.}
\label{fig: luna_corner_1mode}

\end{figure*}

Below are the corner plots showing the posterior chains from the \texttt{LUNA} planet-moon photodynamical model sampled with \texttt{MultiNest}. Figure \ref{fig: luna_corner} shows the full posterior, while Figure \ref{fig: luna_corner_1mode} only shows the best-fit mode based on including the MAP solution.

\bsp	
\label{lastpage}
\end{document}